\documentclass[aps,prd,twocolumn,groupedaddress]{revtex4-2}
\usepackage{blindtext}
\usepackage{amsmath}
\usepackage[symbol]{footmisc} 
\usepackage{color, colortbl}
\usepackage{amsmath} 
\usepackage[utf8]{inputenc}
\usepackage[dvipsnames]{xcolor}
\usepackage[T1]{fontenc}
\usepackage[utf8]{inputenc}
\usepackage{epsfig,latexsym}
\usepackage{adjustbox}
\usepackage{tensor}
\usepackage{graphicx}
\usepackage{subcaption}
\usepackage{verbatim}
\usepackage{mathrsfs}
\usepackage{amssymb}
\usepackage{multirow}
\usepackage{epsfig}
\usepackage{color,colordvi}
\usepackage{appendix}
\usepackage{slashed}
\usepackage{array}
\usepackage[flushleft]{threeparttable}
\usepackage{cancel}
\usepackage{tabularx}
\usepackage{float}
\usepackage[font=footnotesize, justification=centering]{caption}
\usepackage{epsf}
\usepackage{amsmath}
\usepackage{adjustbox}
\usepackage{tikz-feynhand}
\usetikzlibrary{shapes,shadows,arrows}
\usetikzlibrary{arrows}
\tikzset{arrow data/.style 2 args={%
 decoration={%
         markings,
         mark=at position #1 with \arrow{#2}},
         postaction=decorate}
      }%
\usetikzlibrary{svg.path} 
\usetikzlibrary{patterns}
\usepackage{physics}
\usepackage{MnSymbol}
\usepackage{rotating}
\usepackage{multirow}
\usepackage[normalem]{ulem}
\usepackage{tcolorbox}
\usepackage{bbold}
\usepackage{cancel}
\usepackage{ytableau}
\usepackage[vcentermath]{youngtab}
\usepackage{tensor}
\usepackage{autobreak}
\usepackage{tikz}
\usepackage{pgfplots}
\usepackage[figurename=Diagram]{caption}
\pgfplotsset{width=3.3cm,compat=1.8,domain = min:max}
\usepgfplotslibrary{polar}
\usepackage{verbatim}
\usepackage{appendix}
\usetikzlibrary{positioning, fit, calc}
\DeclareMathAlphabet{\mathpzc}{OT1}{pzc}{m}{it}

 \csname
@addtoreset\endcsname{equation}{section}

\newcolumntype{x}[1]{>me{\centering\arraybackslash\hspace{0pt}}p{#1}}
\newcommand{\beq}{\begin{equation}}
\newcommand{\eeq}{\end{equation}}
\renewcommand{\[}{\left[}

\newcommand{\be}{\begin{eqnarray}}
\newcommand{\ee}{\end{eqnarray}}
\newcommand{\bea}{\begin{eqnarray}}
\newcommand{\eea}{\end{eqnarray}}
\newcommand{\bi}{\begin{itemize}}
\newcommand{\ei}{\end{itemize}}
\newcommand{\ben}{\begin{enumerate}}
\newcommand{\een}{\end{enumerate}}
\def\bes{\begin{equation*}}
\def\ees{\end{equation*}}
\def\bead{\begin{aligned}}
\def\eead{\end{aligned}}
\def\bmat{\left(\begin{matrix}}
\def\emat{\end{matrix}\right)}

\def\centerarc[#1](#2)(#3:#4:#5);%
    {
    \draw[#1]([shift=(#3:#5)]#2) arc (#3:#4:#5);
    }

\definecolor{dred}{HTML}{D95F02}
\colorlet{red}{white!15!dred}
\definecolor{darkgreen}{HTML}{1B9E77}
\definecolor{lightgray}{gray}{0.90}

\definecolor{Ecolor}{RGB}{106,157,235}
\definecolor{lightgray}{RGB}{220,220,220}

\definecolor{Gray}{gray}{0.95}

\usepackage{xparse}
\usepackage{tikz}
\usetikzlibrary{calc,matrix,patterns,shadings,backgrounds}
\pgfdeclarelayer{myback}
\pgfsetlayers{myback,background,main}

\tikzset{myfillcolor/.style = {draw,fill=#1}}%

\NewDocumentCommand{\highlight}{O{blue!40} m m}{%
\draw[myfillcolor=#1] (#2.north west)rectangle (#3.south east);
}

\NewDocumentCommand{\vshade}{O{blue!40} O{white} m m}{%
\draw[bottom color =#1,top color=#2] (#3.north west)rectangle (#4.south east);
}

\NewDocumentCommand{\oshade}{O{blue!40} O{white} m m}{%
\draw[right color =#1,left color=#2] (#3.north west)rectangle (#4.south east);
}

\NewDocumentCommand{\inshade}{O{blue!40} O{white} m m}{%
\draw[inner color =#1,outer color=#2] (#3.north west)rectangle (#4.south east);
}

\NewDocumentCommand{\fillpattern}{O{north west lines} O{blue!50} m m}{\draw[pattern=#1, pattern color=#2] (#3.north west)rectangle (#4.south east);
}

\NewDocumentCommand{\pt}{O{north west lines}}{\draw[pattern=north west lines, pattern color=red] (0,0) rectangle (0.3,0.3)}

\begin{document}
\title{\boldmath  Automated Ring-Diagram Framework for Classifying CP Invariants}

\author{Neda Darvishi$^{a}$}
\author{Yining Wang$^{b,c}$}
\author{Jiang-Hao Yu$^{b,c,d,e,f}$}

\affiliation{$^a$Department of Physics, Royal Holloway, University of London, Egham, Surrey, TW20 0EX, United Kingdom}
\email{neda.darvishi@rhul.ac.uk}
\affiliation{$^b$CAS Key Laboratory of Theoretical Physics, Institute of Theoretical Physics, Chinese Academy of Sciences, Beijing 100190, China}
\affiliation{$^c$School of Physical Sciences, University of Chinese Academy of Sciences, Beijing 100049, P. R. China}
\affiliation{$^d$Center for High Energy Physics, Peking University, Beijing 100871, China}
\affiliation{$^e$School of Fundamental Physics and Mathematical Sciences, Hangzhou Institute for Advanced Study, UCAS, Hangzhou 310024, China}
\affiliation{$^f$International Centre for Theoretical Physics Asia-Pacific, Beijing/Hangzhou, China}
\email{wangyining@itp.ac.cn}
\email{jhyu@itp.ac.cn}

\begin{abstract}
\noindent
In this study, we introduce a transformative, automated framework for classifying basis invariants in generic field theories. Utilising a novel ring-diagram methodology accompanied by the well-known Cayley-Hamilton theorem, our approach uniquely enables the identification of basic invariants and their CP-property characterisation. Critically, our framework also unveils previously concealed attributes of established techniques reliant on the Hilbert-Poincaré series and its associated Plethystic logarithm. This paradigm shift has broad implications for the deeper understanding and more accurate classification of CP invariants in generic field theories.
\end{abstract}

\maketitle
\flushbottom

\section{Introduction}
A plethora of experiments and observations signal the emergence of new physics (NP) beyond the Standard Model (BSM), from neutrino masses to matter-antimatter asymmetry and dark matter~\cite{AndreiDSakharov1991,Zwicky:1933gu,QUEST-DMC:2023nug}. Within this context, the existing CP violation (CPV) in the Standard Model (SM) is insufficient to account for the observed matter-antimatter asymmetry of the Universe~\cite{AndreiDSakharov1991}.
Additionally, there is a crucial intersection between CPV and Electric Dipole Moments (EDMs), which violate both parity and time-reversal symmetries~\cite{Dunietz:1985uy,Wu:1985ea,Bernabeu:1986fc,Jenkins:2009dy}. This makes EDMs a direct probe for new sources of CPV. Therefore, any measurable EDM that significantly deviates from SM predictions would strongly indicate NP, including new mechanisms for CPV.
Our research introduces a new method for identifying CP basis invariants, which not only advances our understanding of CPV but also has important implications for interpreting future EDM measurements.  
For a consistent understanding of CPV, it is necessary to use rephasing invariants methods. These are specific mathematical quantities that remain constant under changes to the phase of complex numbers. They help distinguish between physically meaningful parameters and those that are simply artefacts of mathematical representations. The most well-known rephasing invariant is the Jarlskog invariant~\cite{Jarlskog:1985ht,Jarlskog:1985cw}, that is proportional to
\be
{J}^- & \propto (m_t^2-m_c^2)(m_t^2-m_u^2)(m_c^2-m_u^2) \nonumber \\
&\times (m_b^2-m_s^2)(m_b^2-m_d^2)(m_s^2-m_d^2)\, \mathcal{J}\,,
\ee 
where $\mathcal{J}$ can be given in terms of CKM matrix~\cite{Kobayashi:1973fv}
\bea
\mathcal{J} = {\rm Im}\, \Big( V_{us} V_{cb} V_{ub}^\star V_{cs}^\star \Big)\,. \nonumber
\eea
The invariant ${J}^-$ also can be re-written as
\bea
{J}^- = 3\, {\rm Im}\, \mathrm{Det} \Big[ Y_u{Y_u}^{\dagger}\,, Y_d{Y_d}^{\dagger} \Big]\,, \nonumber
\eea
that ensures the theoretical predictions are basis-independent, providing a consistent and reliable means to compare theory with experimental data.

Recent research has leveraged the SM Effective Field Theory (SMEFT) to generalize various NP effects, including those associated with CPV~\cite{Weinberg:1979sa,Buchmuller:1985jz,Braaten1990PRL,Braaten1990PRD,Grzadkowski:2010es,Lehman:2014jma,Henning:2015alf,Li:2020gnx,Murphy:2020rsh,Li:2020xlh,Liao:2020jmn,Liao:2016hru}. However, SMEFT introduces multiple higher-dimensional CP-violating operators, making the task of classifying basis invariants challenging. Tools like the Hilbert-Poincaré series and its Plethystic logarithm (PL) offer some aid toward identifying the number of invariants~\cite{Pouliot_1999,Benvenuti:2006qr,Feng:2007ur,Jenkins:2009dy,Hanany:2014dia,Lehman:2015via,Lehman:2015coa,Henning:2015daa,Henning:2015alf,Jenkins:2009dy,Hanany:2010vu,Wang:2021wdq,Yu:2021cco,Bonnefoy:2021tbt,Yu:2022ttm}. However, the current methods fall short in finding a "primary invariant", especially when it comes to clearly identifying CP-even and CP-odd invariants which are essential for a complete understanding of CPV. 

In this paper, we present an automatic and novel methodology for explicitly classifying CP basis invariants. We utilise fundamental blocks, constructed as orthogonal trivial singlets via newly established Ring-diagrams, and employ the Cayley-Hamilton theorem for precise identification. This paves the way for defining both basic and joint invariants, as well as their CP properties. Our contributions are threefold:
\begin{itemize}
\item We introduce a universal method to classify invariants in diverse theories.
\item We delineate an automatic mechanism for differentiating between CP-odd and -even basis invariants.
\item We elucidate the limitations and hidden aspects of Hilbert series and its PL, thus gaining new structural insights.
\end{itemize}
These fill the existing gap in research, offering a clearer and more complete understanding of CPV in particle physics and cosmology. 

In this paper, we first present our approach within the context of the SM. Subsequently, we outline a more general methodology and illustrate its application using the seesaw SM Effective Field Theory ($\nu$SMEFT) with operators of dimensions 5 and 6.

\section{Identifying Invariants in the SM}

Let us start with the SM Yukawa Lagrangian with up-type and down-type quarks
\bea
\mathcal{L} \supset Y_u \, \overline Q_{{\sf L},i} \widetilde{H}  u_{{\sf R},j}+ Y_d \, \overline Q_{{\sf L},i} ~{H}  d_{{\sf R},j}+\text{h.c.}\ ,
\label{SML}
\eea
where the fundamental quantities in this Lagrangian are the flavor matrices from which the masses and mixing angles
are derived through diagonalisation. Despite being basis-dependent, one can implement unitary transformations on quark and lepton fields. Under chiral flavour transformations, Yukawa couplings evolve as:
\begin{align}
Y_u & \rightarrow U(3)_Q \,Y_u\, {U(3)_u}^{\dagger}, 
\label{Qu-trans}\\
Y_d & \rightarrow U(3)_Q \,Y_d\, {U(3)_d}^{\dagger},
\label{Q-trans}
\end{align}
with $U(3)_Q$ and $U(3)_{u,d}$ being the unitary transformations on the quark doublet $Q_{{\sf L}}$ and singlet fields $u_{{\sf R}},\,u_{{\sf R}}$, remaining the Lagrangian invariant. However, one cannot directly compare the predictions of the mass matrices with their corresponding experimental
values, since these are basis-dependent, whereas observable quantities must be independent of the change of basis. This approach requires cancelling the impact of flavour transformations by parametrising observables in a basis-independent manner, utilising invariant quantities. In this context, the obvious cancellation of $U(3)_{u,d}$ is the combinations
$U \equiv Y_uY^\dagger_{u}$ and $D \equiv Y_d Y^\dagger_{d}$ which both transform as
\begin{align}
U & \rightarrow U(3)_Q \,U\, {U(3)_Q}^{\dagger}, \\
D & \rightarrow U(3)_Q \,D\, {U(3)_Q}^{\dagger}.
\label{UD-trans}
\end{align}
Here, the invariants are trace operations on $U$ and $D$. By exploring all possible combinations of the fundamental blocks $U$ and $D$, we can derive the basic quark invariants. This process is guided by the Cayley–Hamilton theorem, stipulating that the $n$th power of a $n \times n$ matrix $A$ can be expressed in terms of powers less than $n$~\cite{Jenkins:2009dy}.
Thus, the SM invariants $J_{nm}$ (with $n$ and $m$ being the orders of $U$ and $D$, respectively) are thus given by:
\begin{subequations}
\label{eq:SMinv}
\begin{align}
J_{10} &= \Tr(U), \\
J_{01} &= \Tr(D), \\
J_{20} &= \Tr(U^2), \\
J_{02} &= \Tr(D^2), \\
J_{30} &= \Tr(U^3), \\
J_{03} &= \Tr(D^3), \\
J_{11} &= \Tr(U D) = \Tr(D U), \\
J_{21} &= \Tr(U^2 D) = \Tr(D U^2), \\
J_{12} &= \Tr(U D^2) = \Tr(D^2 U), \\
J_{22} &= \Tr(U^2 D^2).
\end{align}
\end{subequations}
Additionally, from the combination of higher-order invariants
\begin{align}
J_{33}&=\Tr\left(U^2 D^2 U D \right),
\\
{J_{33}}'&=\Tr\left(D^2 U^2 D U\right),
\end{align}
where ${\rm CP} \, (J_{33})\to {J_{33}}' $, a CP-odd invariant can be obtained in the following form
\begin{align}
J^- &=  (J_{33})-{(J_{33})}'.
\label{J33m}
\end{align}
Hence, the CPV can be parametrised as
\begin{eqnarray}
J^-&\equiv & {\rm Im}\, \mathrm{Tr} \left[ U, D \right]^3=3 \, {\rm Im}\, \mathrm{Det} \left[ U, D \right].
\label{J4full}
\end{eqnarray}
This is the invariant that appears in the parametrization of the CKM Matrix relevant to CPV in the renormalizable theory of weak interaction~\cite{Cabibbo:1963yz,Kobayashi:1973fv,Wolfenstein:1983yz,Chau:1984fp,Greenberg:1985mr,Braaten1990PRL,Braaten1990PRD}.

Note that CP-even combinations i.e.
\begin{align}
J^+ &= (J_{33})+{(J_{33})}',
\label{J33p}
\end{align}
can be rewritten in terms of the lower primary invariants, thus this is disregarded as a basic invariant. 

Now, we introduce an alternative procedure that methodically classifies invariants in the SM and extends further. This classification is based on a newly established ring-diagram that is accompanied by Cayley--Hamilton theorem and equivalently by ring-set (petals) enabling us to identify basic invariants. Particularly, we show our approach automatically distinguishes basic invariants as well as their CP properties.

To begin, we designate three rings corresponding to the flavor transformations $U(3)_Q, \,{U(3)_u}$, and ${U(3)_d}$ of the Yukawa matrices, as illustrated in Table \ref{tab:tab1}.
\begin{table}[H]
	\centering
	\begin{tabular}{c|c|c|c}
		& $U(3)_Q$ & $U(3)_u$ & $U(3)_d$ \\\hline
		$Y_u$ &$\mathbf{3}$ & $ \mathbf{\overline{3}}$ & $\mathbf{1}$  \\[0.1cm]
		$Y_d$ &$\mathbf{3}$ & $ \mathbf{1}$ & $\mathbf{\overline{3}}$  \\[0.1cm]
	\end{tabular}
	\caption{\it Flavor transformation of the Yukawa matrices.}
	\label{tab:tab1}
\end{table} 
In this configuration, as delineated in Diagram~\ref{D1}, $Y_u$ and $Y_u^\dagger$ are positioned within the ${U(3)_u}$-ring, while $Y_d$ and $Y_d^\dagger$ are located within the ${U(3)_d}$-ring. The $U(3)_Q$-ring acts as a connecting hub between these two rings. The diagrams below succinctly represent both the detailed and simplified versions of this structure.
\setcounter{figure}{0}
\renewcommand{\figurename}{Diagram}
\begin{figure}[t]
\begin{center}
\includegraphics[width=0.4\textwidth]{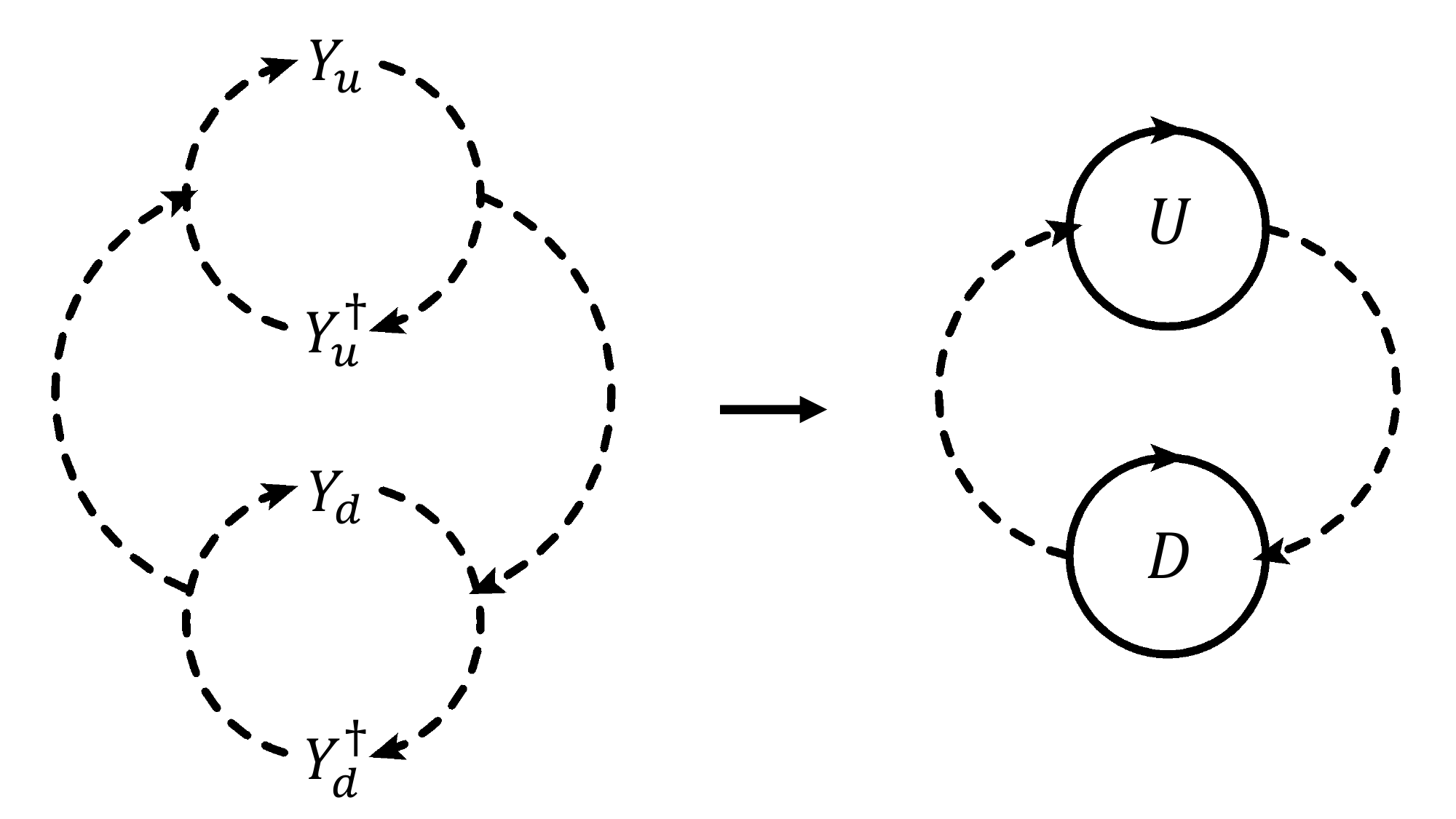}
\end{center}
\caption{Ring-diagram for the SM.}
\label{D1}
\end{figure}

Now, we construct the three building blocks pertinent to these three rings with only one passage through them, as
\ytableausetup{smalltableaux}
\bea
&M_u=\text{diag}(U,\,0) \equiv  \hspace{0.2cm}{\raisebox{-0.7em}{\begin{tikzpicture}
\def \radius {0.4cm}
  \node[draw, circle,style={thick},-latex] {$U$};
  \draw[-latex,->,thick] ({180}:\radius);
\end{tikzpicture}}}\,,
\quad M_d=\text{diag}(0,\,D)\equiv  \hspace{0.2cm}{\raisebox{-0.7em}{\begin{tikzpicture}
\def \radius {0.4cm}
  \node[draw, circle,style={thick},-latex] {$D$};
  \draw[-latex,->,thick] ({180}:\radius);
\end{tikzpicture}}}\,
\nonumber \\
& M_{ud}^\pm \equiv \begin{pmatrix}
 U D & 0\\
 0 &  \pm \,D U \\
\end{pmatrix}~ \equiv \raisebox{-0.5em}{ 
\begin{tikzpicture}[transform shape,line width=0.9pt]
\node (A) at (0,0) {$U$}; 
\node (B) at (1.2,0) {$D$}; 
\path[draw] 
(A) edge[bend right=-40,black,dashed,->] (B)
(B) edge[bend right=-40,black,-latex,dashed,->] (A);
\end{tikzpicture}} \,.
 \nonumber \\
 \label{blocks-SM}
\eea
Here, the trace operations over the rings form what we term as "petals". The petal formations are depicted below, illustrating the trace operations and their corresponding mathematical expressions:
\begin{subequations}
\begin{eqnarray}
\ytableausetup{boxsize=1.6em}\ytableausetup{smalltableaux}
\Tr {\raisebox{-0.7em}{\begin{tikzpicture}
\def \radius {0.4cm}
  \node[draw, circle,style={thick},-latex] {$U$};
  \draw[-latex,->,thick] ({180}:\radius);
\end{tikzpicture}}~} & :=&  \begin{tikzpicture}
   \begin{polaraxis}[grid=none, axis lines=none]
     	\addplot+[mark=none,domain=0:90,samples=50,color=red,style={ultra thick}] 
		{sin(2*x)}; 
   \end{polaraxis}
 \end{tikzpicture}  
= \Tr (U),  
\\
 \Tr{\raisebox{-0.7em}{\begin{tikzpicture}
\def \radius {0.4cm}
  \node[draw, circle,style={thick},-latex] {$D$};
  \draw[-latex,->,thick] ({180}:\radius);
\end{tikzpicture}}~}& :=& \hspace{0.0em}{ \begin{tikzpicture}
   \begin{polaraxis}[grid=none, axis lines=none]
     	\addplot+[mark=none,domain=0:90,samples=50,color=blue,style={ultra thick}] 
		{sin(2*x)}; 
   \end{polaraxis}
 \end{tikzpicture}}=  \Tr(D), 
\\
\Tr {\hspace{-0.2em}{\raisebox{-0.5em}{ 
\begin{tikzpicture}[transform shape,line width=0.9pt]
\node (A) at (0,0) {$U$}; 
\node (B) at (1.2,0) {$D$}; 
\path[draw] 
(A) edge[bend right=-40,black,dashed,->] (B)
(B) edge[bend right=-40,black,-latex,dashed,->] (A);
\end{tikzpicture}}} }^+ &:=&  \hspace{0.0em}{\raisebox{-1.5
em}{\begin{tikzpicture}
   \begin{polaraxis}[grid=none, axis lines=none]
     	\addplot+[mark=none,domain=0:90,samples=50,color=red,style={ultra thick}] 
		{sin(2*x)}; 
   \end{polaraxis}
    \begin{polaraxis}[grid=none, axis lines=none]
     	\addplot+[mark=none,domain=0:90,samples=50,color=blue,style={ultra thick}] 
		{-sin(2*x)}; 
   \end{polaraxis}
 \end{tikzpicture}}}
= \Tr (U D),  
\\
 \Tr {\hspace{-0.2em}{\raisebox{-0.5em}{ 
\begin{tikzpicture}[transform shape,line width=0.9pt]
\node (A) at (0,0) {$U$}; 
\node (B) at (1.2,0) {$D$}; 
\path[draw] 
(A) edge[bend right=-40,black,dashed,->] (B)
(B) edge[bend right=-40,black,-latex,dashed,->] (A);
\end{tikzpicture}}}}^-
& :=& \hspace{-0.2em}{\raisebox{-1.5
em}{\begin{tikzpicture}
   \begin{polaraxis}[grid=none, axis lines=none]
     	\addplot+[mark=none,domain=0:90,samples=50,color=red,style={ultra thick}] 
		{sin(2*x)}; 
   \end{polaraxis}
    \begin{polaraxis}[grid=none, axis lines=none]
     	\addplot+[mark=none,domain=0:90,samples=50,color=blue,style={ultra thick}] 
		{-sin(2*x)}; 
   \end{polaraxis}
 \end{tikzpicture}}}
= 0.
\label{read-ring1}
\end{eqnarray}
\end{subequations}
These petals effectively represent the component matrices $Y_u$, $Y_u^\dagger$, $Y_d$, and $Y_d^\dagger$, each a $3 \times 3$ matrix, transforming according to U(3)$_u$ and U(3)$_d$, respectively. Utilising the foundational blocks and rings defined above, we can systematically organise higher-order invariants as shown in the following tensor product equation:
\begin{equation}
J_{{u}^{x_1} {d}^{x_2} {(ud)}^{x_n}} = \Tr \left({M_{u}}^{x_1} \otimes  {M_d}^{x_2} \otimes  {M_{ud}}^{ x_n} \right).
\end{equation}
In an equivalent manner, petal sets can be effectively utilised in our approach. Considering that $U$ and $D$ are $3\times3$ matrices, rings can accommodate up to three passages over them. While combinations between rings are allowed, the number of petals in each combination should not exceed three. Therefore, when constructing higher-order invariants using petals, one can repeat up to two out of the three petals of the same color. This mandates that the third petal of the same color must follow a petal of a different hue. To simplify the petal set, two adjacent petals can be replaced with a dual-layer petal. This approach can also be applied to the repetition of multiple petals, but only twice i.e. \ytableausetup {boxsize=1.5em}\bea
 \hspace{-0.1cm}{\raisebox{-1.3em}{ \begin{tikzpicture}
\begin{polaraxis}[grid=none, axis lines=none]
\addplot+[mark=none,domain=0:270,samples=50,color=blue,style={ultra thick} ] 
		{-sin(2*x)};  
   \end{polaraxis}
 \begin{polaraxis}[grid=none, axis lines=none]
\addplot+[mark=none,domain=0:270,samples=50,color=red,style={ultra thick} ] 
		{+sin(2*x)};  
   \end{polaraxis}
      \begin{polaraxis}[grid=none, axis lines=none]
\addplot+[mark=none,domain=0:90,samples=50,color=red,style={ultra thick}] 
		{sin(2*x)};  
   \end{polaraxis}
    \begin{polaraxis}[grid=none, axis lines=none]
\addplot+[mark=none,domain=90:180,samples=50,color=blue,style={ultra thick}] 
		{sin(2*x)};  
   \end{polaraxis} 
    \end{tikzpicture}}} ~\equiv~ \hspace{-0.6cm}{\raisebox{-3.2em}{ 
\begin{tikzpicture}
\hspace{0.7cm}{\raisebox{1.8em}{  \begin{polaraxis}[grid=none,  axis lines=none]
     	\addplot+[mark=none,domain=0:90,samples=50,color=red,style={ultra thick}] 
		{ sin(2*x)};
   \end{polaraxis}}}
   \hspace{-0.3cm}{\raisebox{1.1em}{  \begin{polaraxis}[width=3.9cm,grid=none,  axis lines=none]
     	\addplot+[mark=none,domain=0:90,samples=50,color=red,style={ultra thick}] 
		{ sin(2*x)};
   \end{polaraxis}}}
   \hspace{-0.42cm}{\raisebox{0em}{ 
\hspace{0.7cm}{\raisebox{1.8em}{  \begin{polaraxis}[grid=none,  axis lines=none]
     	\addplot+[mark=none,domain=0:90,samples=50,color=blue,style={ultra thick}] 
		{ -sin(2*x)};
   \end{polaraxis}}}
   \hspace{-0.3cm}{\raisebox{1.1em}{  \begin{polaraxis}[width=3.9cm,grid=none,  axis lines=none]
     	\addplot+[mark=none,domain=0:90,samples=50,color=blue,style={ultra thick}] 
		{-sin(2*x)};
   \end{polaraxis}}}}}
 \end{tikzpicture}}}
 \nonumber
    \eea
After this simplification one can translate them to the relations by taking an arbitrary starting point in rings and read a whole cycle clockwise $\pm$ anti-clockwise  i.e. 
$$J^\pm \equiv (\circlearrowright +\searrow\swarrow \nwarrow \cdots) \pm (\circlearrowleft+\searrow\nearrow\nwarrow \cdots).$$

Therefore, using the fundamental blocks~\eqref{blocks-SM} the SM invariants can be recovered as
{\allowdisplaybreaks\ytableausetup{smalltableaux}
\bea
J_{u^1}&=&\Tr( M_u):= \begin{tikzpicture}
   \begin{polaraxis}[grid=none, axis lines=none]
     	\addplot+[mark=none,domain=0:90,samples=50,color=red,style={ultra thick}] 
		{sin(2*x)}; 
   \end{polaraxis}
 \end{tikzpicture}  
= \Tr (U) 
 \nonumber \\
J_{d^1}&=&\Tr(M_d):= \hspace{0.0em}{ \begin{tikzpicture}
   \begin{polaraxis}[grid=none, axis lines=none]
     	\addplot+[mark=none,domain=0:90,samples=50,color=blue,style={ultra thick}] 
		{sin(2*x)}; 
   \end{polaraxis}
 \end{tikzpicture}}=  \Tr (D)
 \nonumber \\
J_{ud^1}&=&\Tr (M_{ud}):=  \hspace{0.0em}{\raisebox{-1.5
em}{\begin{tikzpicture}
   \begin{polaraxis}[grid=none, axis lines=none]
     	\addplot+[mark=none,domain=0:90,samples=50,color=red,style={ultra thick}] 
		{sin(2*x)}; 
   \end{polaraxis}
    \begin{polaraxis}[grid=none, axis lines=none]
     	\addplot+[mark=none,domain=0:90,samples=50,color=blue,style={ultra thick}] 
		{-sin(2*x)}; 
   \end{polaraxis}
 \end{tikzpicture}}}
= \Tr (U D) 
\nonumber \\
J_{u^2}&=&\Tr(M_u^{ 2}):= \hspace{0.0em}{\raisebox{-1.5
em}{\begin{tikzpicture}
   \begin{polaraxis}[grid=none, axis lines=none]
     	\addplot+[mark=none,domain=0:90,samples=50,color=red,style={ultra thick}] 
		{sin(2*x)}; 
   \end{polaraxis}
    \begin{polaraxis}[grid=none, axis lines=none]
     	\addplot+[mark=none,domain=0:90,samples=50,color=red,style={ultra thick}] 
		{-sin(2*x)}; 
   \end{polaraxis}
 \end{tikzpicture}}}= \Tr (U^2) \nonumber \\
J_{d^2} &=&\Tr(M_d^{2}):=  \hspace{0.0em}{\raisebox{-1.5
em}{\begin{tikzpicture}
   \begin{polaraxis}[grid=none, axis lines=none]
     	\addplot+[mark=none,domain=0:90,samples=50,color=blue,style={ultra thick}] 
		{sin(2*x)}; 
   \end{polaraxis}
    \begin{polaraxis}[grid=none, axis lines=none]
     	\addplot+[mark=none,domain=0:90,samples=50,color=blue,style={ultra thick}] 
		{-sin(2*x)}; 
   \end{polaraxis}
 \end{tikzpicture}}}=\Tr (D^2)
 \nonumber \\
J_{u^1{ud}^1}&=&\Tr(M_u \otimes M_{ud} ):= \hspace{0cm}{\raisebox{-1.1em}{ 
\begin{tikzpicture}
 \begin{polaraxis}[grid=none,  axis lines=none]
     	\addplot+[mark=none,domain=0:60,samples=50,color=blue,style={ultra thick}] 
		{-sin(3*x)}; 
   \end{polaraxis}
   \begin{polaraxis}[grid=none,  axis lines=none]
     	\addplot+[mark=none,domain=61:180,samples=50,color=red,style={ultra thick}] 
		{-sin(3*x)}; 
		\end{polaraxis}
 \end{tikzpicture}}}= \Tr (U^2 D) 
\nonumber \\
J_{{ud}^1d^1}&=&\Tr(M_d \otimes M_{ud}):=\hspace{0cm}{\raisebox{-1.1em}{ 
\begin{tikzpicture}
 \begin{polaraxis}[grid=none,  axis lines=none]
     	\addplot+[mark=none,domain=60:120,samples=50,color=red,style={ultra thick}] 
		{-sin(3*x)}; 
   \end{polaraxis}
   \begin{polaraxis}[grid=none,  axis lines=none]
     	\addplot+[mark=none,domain=0:60,samples=50,color=blue,style={ultra thick}] 
		{-sin(3*x)}; 
		\end{polaraxis}
		 \begin{polaraxis}[grid=none,  axis lines=none]
     	\addplot+[mark=none,domain=120:180,samples=50,color=blue,style={ultra thick}] 
		{-sin(3*x)}; 
		\end{polaraxis}
 \end{tikzpicture}}}= \Tr( D^2U)
 \nonumber \\
J_{u^3}&=&\Tr(M_u^{ 3}):=\hspace{0cm}{\raisebox{-1.1em}{ 
\begin{tikzpicture}
 \begin{polaraxis}[grid=none,  axis lines=none]
     	\addplot+[mark=none,domain=0:180,samples=50,color=red,style={ultra thick}] 
		{-sin(3*x)}; 
   \end{polaraxis}
 \end{tikzpicture}}}= \Tr (U^3)
  \nonumber \\
J_{d^3} &=&\Tr(M_d^{3}) :=   \hspace{0cm}{\raisebox{-1.1em}{ 
\begin{tikzpicture}
 \begin{polaraxis}[grid=none,  axis lines=none]
     	\addplot+[mark=none,domain=0:180,samples=50,color=blue,style={ultra thick}] 
		{-sin(3*x)}; 
   \end{polaraxis}
\end{tikzpicture}}}= \Tr (D^3)
 \nonumber \\
J_{ud^2}&=&\Tr(M_{ud}^{ 2}):=   \hspace{0.0cm}{\raisebox{-1.4em}{ \begin{tikzpicture}
\begin{polaraxis}[grid=none, axis lines=none]
\addplot+[mark=none,domain=0:180,samples=50,color=red,style={ultra thick}] 
		{+sin(2*x)};  
   \end{polaraxis}
 \begin{polaraxis}[grid=none, axis lines=none]
\addplot+[mark=none,domain=180:360,samples=50,color=blue,style={ultra thick}] 
		{+sin(2*x)};  
   \end{polaraxis}
    \end{tikzpicture}}}= \Tr ((U D)^2) 
\nonumber
\label{Inv-SM}
\eea}
and an order-6 non-reducible CP-odd invariant can be deduced out of
\begin{eqnarray}
{J^-}_{ud^3}&=&\Tr(M_{ud}^{ 3}):= {\raisebox{-1.5em}{\begin{tikzpicture}
 \begin{polaraxis}[grid=none,  axis lines=none]
     	\addplot+[mark=none,domain=0:270,samples=50,color=red,style={ultra thick}] 
		{-sin(3*x)}; 
   \end{polaraxis}
   \begin{polaraxis}[grid=none,  axis lines=none]
     	\addplot+[mark=none,domain=0:270,samples=50,color=blue,style={ultra thick}] 
		{sin(3*x)}; 
   \end{polaraxis}
 \end{tikzpicture}}} 
 \nonumber\\
 &=& \Big[\Tr (U^2 D^2 U D)\,-\,\Tr (D^2 U^2 D U)\Big].
  \nonumber 
 \label{Inv-SM-CPO}
 \end{eqnarray}

In the subsequent section, we expand this formalism to identify potential invariants in a generic scenario, underscoring the versatility and applicability of our approach.

\section{Generic formalism for identifying invariants}
In many complex systems, it is useful to build and understand the relation between invariants. In this letter, we introduce the so-called Ring-diagram, which comprises sets of interconnected rings, each adhering to a specific symmetry constraint. 

\begin{figure}[t]
\begin{center}
\includegraphics[width=0.17\textwidth]{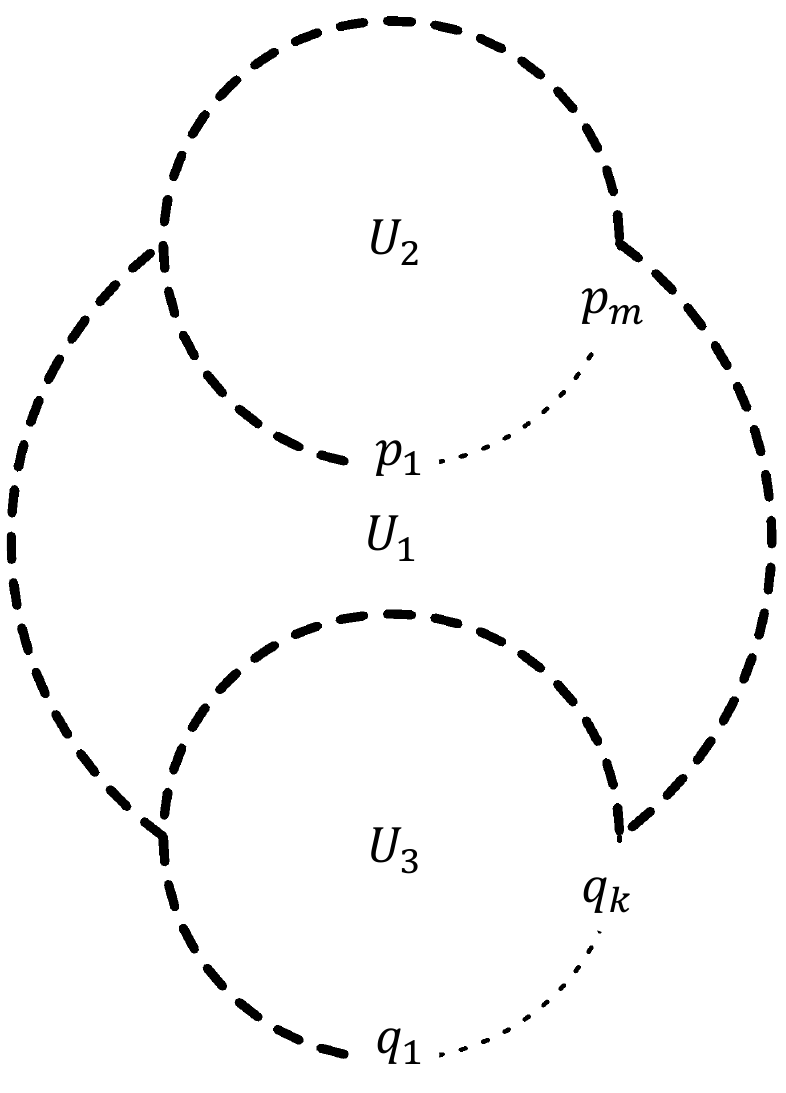}
\end{center}
\caption{The generic Ring-diagram.}
\label{R1}
\end{figure}

Let us begin with a generic example involving $n = m + k$ main elements $p_1, \cdots, p_m$ and $q_1, \cdots, q_k$, where the $p$ elements rotate with $U_1$ and $U_2$, while the $q$ elements rotate with $U_1$ and $U_3$. To represent these interactions, we introduce three rings: the first ring labelled with $U_2$ representing rotations of $p$ elements with $U_2$, the second ring labelled with $U_3$ for rotations of $q$ elements with $U_3$, and the third ring connecting all rings relevant to mutual $U_1$. This structure is depicted in Ring-diagram~\ref{R1}, where the number of rings can increase due to the presence of more rotations with $U_i$. Here, clusters of interlinked rings form building blocks that are characterized by symmetries.

\begin{figure}[t]
\begin{center}
\begin{subfigure}{0.2\textwidth}
  \includegraphics[width=\linewidth]{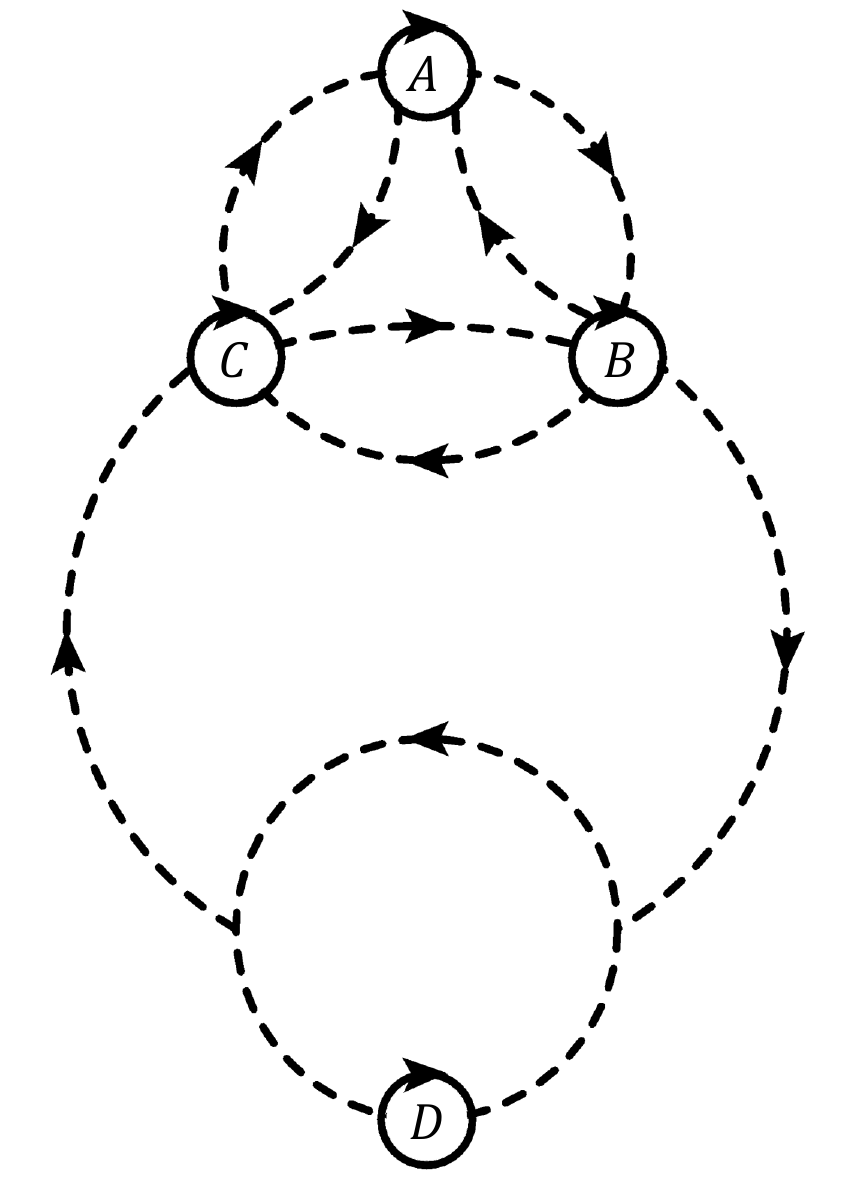}
  \caption{ }  
  \label{subfig:RD2}
\end{subfigure}\quad
\begin{subfigure}{0.19\textwidth}
  \includegraphics[width=\linewidth]{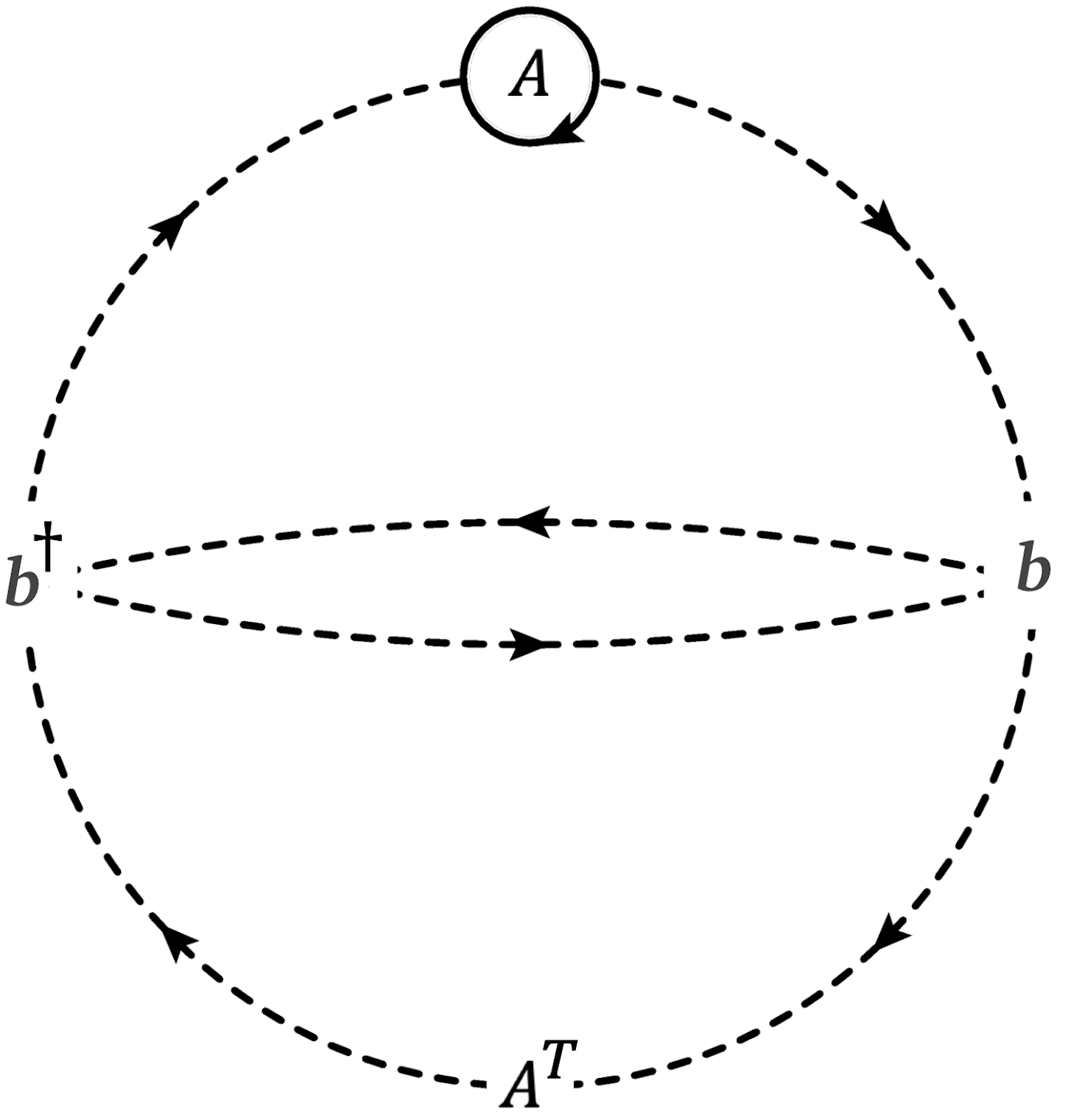}
  \caption{ }  
  \label{subfig:RD6}
\end{subfigure}
\end{center}
\caption{Ring-diagrams for (a) non-involving blocks in other rings, and (b) involving blocks.}
\label{RD}
\end{figure}

The inputs to the rings fall into two categories, aiding the construction of possible invariants. The first type involves first-order blocks, easily obtained through combinations of main elements within the $U_i$-ring, where $U_i$ cancels out. These blocks are placed in the relevant ring with arbitrary sequences and are denoted by a small circle ``$\circlearrowright$''. Blocks are connected with dashed cycles, represented by $\begin{tikzpicture}[transform shape,line width=1pt] \node (A) at (2,2) {}; \node (B) at (3,2) {}; \path[draw] (A) edge[bend right=-30,black,dashed,->] (B) (B) edge[bend right=-30,black,-latex,dashed,->] (A); \end{tikzpicture}$.
The second input type pertains to elements where cancellation of $U_i$ in their rotations involves combinations directly used in building blocks in other rings. These elements remain unchanged in the rings and connect to other elements or blocks through single-way dashed curves $\begin{tikzpicture}[transform shape,line width=1pt] \node (A) at (2,2) {}; \node (B) at (3,2) {}; \path[draw] (A) edge[bend right=-30,black,dashed,->] (B); \end{tikzpicture}$.
In this case, a cycle of dashed lines forms a building block, with only one turn passing through the connection lines as 
$$\begin{adjustbox}{max height=0.07\textheight, max width=0.4\textwidth,center}\begin{tikzpicture}[transform shape,line width=1pt]
\node (A) at (1,1) { x}; 
\node (B) at (3,1) { x}; 
\path[draw] 
(A) edge[bend right=-30,black,dashed,->] (B)
(B) edge[bend right=-30,black,dashed,->] (A);
\end{tikzpicture},\,
\begin{tikzpicture}[transform shape,line width=1pt]
\node (A) at (1,1) {x }; 
\node (B) at (2,2) {x }; 
\node (C) at (3,1) { x}; 
\path[draw] 
(A) edge[bend right=-30,black,dashed,->] (B) 
(B) edge[bend right=-30,black,dashed,->] (C)
(C) edge[bend right=-30,black,dashed,->] (A);
\end{tikzpicture}, \,
\begin{tikzpicture}[transform shape,line width=1pt]
\node (A) at (1,1) {x }; 
\node (B) at (2,2) { x}; 
\node (C) at (3,1) {x };
\node (D) at (2,0) { x}; 
\node (E) at (1, 1) {x }; 
\path[draw] 
(A) edge[bend right=-30,black,dashed,->] (B) 
(B) edge[bend right=-30,black,dashed,->] (C)
(C) edge[bend right=-30,black,dashed,->] (D) 
(D) edge[bend right=-30,black,dashed,->] (E);
\end{tikzpicture}\end{adjustbox}\hspace{-0.6in}, \cdots \;.$$
Although the mutual $U_1$ ring allows for $N=N_1\times N_2$ pass through the connection lines with $N_{1,2}$ being the number of blocks in $U_{2,3}$ rings, this restriction helps us to build the lowest possible combinations in the construction of blocks.
Therefore, in a ring-diagram one may have $m_1$ number of first-order blocks tagged by ``$\circlearrowright $'' and $m_2$ building blocks out of dashed cycles. 
 Accordingly, one can construct these building blocks in the form of tensors and produce higher-order invariants. The former blocks can be arranged in sets of $m_1$ orthogonal vectors.
For instance, in ring-diagrams \ref{RD} (a) and (b), we have two different sets $m_1=4$ and $m_1=2, m_2=1$ building blocks, where $m_1$ are first-order blocks tagged with ``$\circlearrowright$'' and $m_2$ are involving building blocks out of dashed cycles.
 
Accordingly, for each diagram the building blocks can be given in the form of matrices. For example, the building blocks pertinent to diagrams~(a) 
\begin{align}
 &M_a \equiv \text{diag} \begin{pmatrix} A & 0 & 0 & 0
\end{pmatrix},\quad 
 M_b \equiv \text{diag} \begin{pmatrix} 0 & B & 0 & 0  \end{pmatrix},
 \nonumber \\
 &M_c \equiv \text{diag} \begin{pmatrix} 0 & 0 & C & 0 \end{pmatrix},\quad  
  M_d \equiv \text{diag} \begin{pmatrix}0 & 0 & 0 & D \end{pmatrix},
\end{align}
and for diag (b) $M_a$ remains same while $M_b$ read as
\begin{align}
& {M_{b}} \equiv \sigma_0 \otimes \begin{pmatrix}
b \,b^{\dagger} & 0 \\
0 & \, b^{\dagger}\,b   \\
\end{pmatrix}~ \equiv \raisebox{-1em}{ 
\begin{tikzpicture}[transform shape,line width=0.9pt]
\node (A) at (0,0) {$b$}; 
\node (B) at (1.2,0) { $b^{\dagger}$ }; 
\path[draw] 
(A) edge[bend right=-40,black,dashed,->] (B)
(B) edge[bend right=-40,black,-latex,dashed,->] (A);
\end{tikzpicture}}\equiv  \hspace{0.2cm}{\raisebox{-0.7em}{\begin{tikzpicture}
\def \radius {0.4cm}
  \node[draw, circle,style={thick},-latex] {$B$};
  \draw[-latex,->,thick] ({180}:\radius);
\end{tikzpicture}}}\,.
\end{align}
For higher-order blocks elements follow dashed cycles between first-order blocks in the following forms for diagrams (a) and (b)
\begin{align}
& M_{ab}^{\pm} \equiv {\sigma_0}\otimes \begin{pmatrix}
\raisebox{-0.9em\begin{tikzpicture}[transform shape,line width=0.6pt]
\node (A) at (0,0) {A }; 
\node (B) at (0.9,0) {B}; 
\path[draw] 
(A) edge[bend right=-40,black,dashed,-> ] (B);
\end{tikzpicture}}
& \,0 \\
0 \,& \pm\raisebox{-0.9em}{\begin{tikzpicture}[transform shape,line width=0.6pt]
\node (A) at (0,0) {A}; 
\node (B) at (0.9,0) {B}; 
\path[draw] 
(B) edge[bend right=-40,black, dashed,->] (A);
\end{tikzpicture}}
\end{pmatrix},
\label{Mab}
\\
& M_{ac}^\pm \equiv   \sigma_0 \otimes \begin{pmatrix}
A \,b\,{A^{\sf T}}^m\,b^{\dagger}& 0\\
0 &\,  \pm b \,{A^{\sf T}}^m\,b^{\dagger}\,A  \\
\end{pmatrix}\equiv \hspace{-1em}{\raisebox{-0.1em}{ 
\raisebox{-2.8em}{ \begin{tikzpicture}[transform shape,line width=1pt]
\node (A) at (0.9,1) {$b^{\dagger}$}; 
\node (B) at (2,2) {$A$}; 
\node (C) at (3,1) {$b$};
\node (D) at (2,0) {${A^{\sf T}}^m$}; 
\node (E) at (1,0.8) {}; 
\path[draw] 
(A) edge[bend right=-30,black,dashed,->] (B) 
(B) edge[bend right=-30,black,dashed,->] (C)
(C) edge[bend right=-30,black,dashed,->] (D) 
(D) edge[bend right=-30,black,dashed,->] (E);
\end{tikzpicture}}}}\,
\nonumber \\
& \equiv \sigma_0 \otimes \begin{pmatrix}
A \,C^m & 0\\
0 &\,  \pm C^m\,A  \\
\end{pmatrix}\equiv \raisebox{-1em}{ 
\begin{tikzpicture}[transform shape,line width=0.9pt]
\node (A) at (0,0) {$A$}; 
\node (B) at (1.2,0) { $C^m$ }; 
\path[draw] 
(A) edge[bend right=-40,black,dashed,->] (B)
(B) edge[bend right=-40,black,-latex,dashed,->] (A);
\end{tikzpicture}}\,,
 \label{blocks-dim-6-2}
\end{align}
with $\sigma_0=\bf{1}_2$ and $C^m=b \,{A^{\sf T}}^m\,b^{\dagger}$. 

The lowest order of invariants is formed based on trace over interlinked rings referred to as petals. These components take the form of orthogonal trivial singlets and are positioned uniquely within the framework of the ring-diagram structure. The properties of the building blocks can be directly represented in terms of petals. One cycle of first-order blocks that are tagged by ``$\circlearrowright $'' in Ring-diagram~\ref{RD} can be translated into a single petal representation. For higher order cycles depending on their $n$ connection lines $n$ petals are required e.g. $n=2$ for ``$\begin{tikzpicture}[transform shape,line width=1pt]
\node (A) at (2,2) { }; 
\node (B) at (3,2) { }; 
\path[draw] 
(A) edge[bend right=-30,black,dashed,->] (B)
(B) edge[bend right=-30,black,-latex,dashed,->] (A);
\end{tikzpicture}$''. This structure from the ring-diagram can be summarised as follows:
{\allowdisplaybreaks
\ytableausetup{boxsize=1.6em}\begin{eqnarray}
\hspace{0cm}{\raisebox{-0.7em}{\begin{tikzpicture}
\def \radius {0.4cm}
  \node[draw, circle,style={thick},-latex] {A};
  \draw[-latex,->,thick] ({180}:\radius);
\end{tikzpicture}}}~:=&~
\raisebox{-0.2em}{\begin{tikzpicture}
   \begin{polaraxis}[grid=none, axis lines=none]
     	\addplot+[mark=none,domain=0:90,samples=50,color=black,style={ultra thick}] 
		{sin(2*x)}; 
   \end{polaraxis}
 \end{tikzpicture}}\;,\nonumber\\
\raisebox{-0.9em}{ \begin{tikzpicture}[transform shape,line width=1pt]
\node (A) at (0,0) {A}; 
\node (B) at (1.5,0) {B}; 
\path[draw] 
(A) edge[bend right=-40,black,dashed,->] (B)
(B) edge[bend right=-40,black,-latex,dashed,->] (A);
\end{tikzpicture}}  
 :=&
\raisebox{-1em}{\begin{tikzpicture}
   \begin{polaraxis}[grid=none, axis lines=none]
     	\addplot+[mark=none,domain=0:90,samples=50,color=black,style={ultra thick}] 
		{sin(2*x)}; 
   \end{polaraxis}
      \begin{polaraxis}[grid=none, axis lines=none]
     	\addplot+[mark=none,domain=0:90,samples=50,color=black,style={ultra thick},style=dashed] 
		{-sin(2*x)}; 
   \end{polaraxis}
 \end{tikzpicture}}\;,
    \nonumber \\ 
\raisebox{-0.8em}{\begin{tikzpicture}[transform shape,line width=1pt]
\node (A) at (1,1) {C }; 
\node (B) at (2,2) {A}; 
\node (C) at (3,1) {D}; 
\path[draw] 
(A) edge[bend right=-30,black,dashed,->] (B) 
(B) edge[bend right=-30,black,dashed,->] (C)
(C) edge[bend right=-30,black,dashed,->] (A);
\end{tikzpicture}}  :=&
\raisebox{-1em}{ \begin{tikzpicture}
   \begin{polaraxis}[grid=none, axis lines=none]
\addplot+[mark=none,domain=0:60,samples=50,color=black,style={ultra thick},style=dotted] 
		{sin(-3*x)};  
   \end{polaraxis}
     \begin{polaraxis}[grid=none, axis lines=none]
\addplot+[mark=none,domain=60:120,samples=50,color=black,style={ultra thick}] 
		{sin(-3*x)};  
   \end{polaraxis}
     \begin{polaraxis}[grid=none, axis lines=none]
\addplot+[mark=none,domain=120:180,samples=50,color=black,style={ultra thick},style=dashdotted] 
		{sin(-3*x)};  
   \end{polaraxis}
 \end{tikzpicture}}\;, \nonumber\\
\raisebox{-2.8em}{ \begin{tikzpicture}[transform shape,line width=1pt]
\node (A) at (1,1) {G}; 
\node (B) at (2,2) {A}; 
\node (C) at (3,1) {E};
\node (D) at (2,0) {F}; 
\node (E) at ( 1, 1) {}; 
\path[draw] 
(A) edge[bend right=-30,black,dashed,->] (B) 
(B) edge[bend right=-30,black,dashed,->] (C)
(C) edge[bend right=-30,black,dashed,->] (D) 
(D) edge[bend right=-30,black,dashed,->] (E);
\end{tikzpicture}} :=&
\raisebox{-1.3em}{ \begin{tikzpicture}
   \begin{polaraxis}[grid=none, axis lines=none]
\addplot+[mark=none,domain=0:90,samples=50,color=black,style={ultra thick}] {sin(2*x)};  
   \end{polaraxis}
    \begin{polaraxis}[grid=none, axis lines=none]
\addplot+[mark=none,domain=90:180,samples=50,color=black,style={ultra thick},style=densely dashed] {sin(2*x)};  
   \end{polaraxis}
    \begin{polaraxis}[grid=none, axis lines=none]
\addplot+[mark=none,domain=180:270,samples=50,color=black,style={ultra thick},style=densely dashdotdotted] {sin(2*x)};  
   \end{polaraxis}
    \begin{polaraxis}[grid=none, axis lines=none]
\addplot+[mark=none,domain=270:360,samples=50,color=black,style={ultra thick},style=loosely dashed] {sin(2*x)};  
   \end{polaraxis}
 \end{tikzpicture}}, \, \cdots\,. 
 \nonumber
 \label{read-ring}
\end{eqnarray}}
Furthermore, the petals related to the blocks are distinguished with different colours/patterns.
Note that these representations always follow the prearranged direction in the ring-diagram. Rings facilitate the construction of higher-order identical invariants as any repetitions even illustriously would be trivial to drop out. This formulation shows the connection between invariants as well as distinguishing their CP properties.

To step forward, the building blocks can be extended up to higher orders equivalently using diagrams or based on the products of the above building blocks
\begin{equation}
J_{{n_a}^{x_1},\,{n_b}^{x_2},\,\cdots,\,{n_{ab}}^{x_n}} = \Tr \left({M_{a}}^{x_1} \otimes  {M_b}^{x_2} \otimes \cdots  \otimes  {M_{ab}}^{ x_n} \right).
\label{J-general}
\end{equation}
To construct invariants using diagrams \eqref{read-ring},  firstly one may take an arbitrary starting point in rings that are assigned to a fundamental block and read a whole cycle according to the possible existing discrete symmetries in the common ring for example in the most cases clockwise $\pm$ anti-clockwise and repeated until identical terms produced e.g.
\bea 
\hspace{-.2cm}{\raisebox{-1.2em}{\begin{tikzpicture}
  \begin{polaraxis}[grid=none,  axis lines=none]
     	\addplot+[mark=none,domain=90:270,samples=50,color=red,style={ultra thick}] 
		{-sin(2*x)}; 
   \end{polaraxis}
   \begin{polaraxis}[grid=none,  axis lines=none]
     	\addplot+[mark=none,domain=0:90,samples=50,color=darkgreen,style={ultra thick}] 
		{-sin(2*x)}; 
		\end{polaraxis}
 \begin{polaraxis}[grid=none,  axis lines=none]
     	\addplot+[mark=none,domain=270:360,samples=50,color=blue,style={ultra thick}] 
		{-sin(2*x)}; 
\end{polaraxis}
 \end{tikzpicture}}}&& =(R\, R\, B\, G + R\, B\,R\, G) \, \pm \,  (R\, G\,B\, R + R\, B\,R\,G)
 \nonumber \\
&&\equiv (R\, R\, B\, G + R\, R\, G\, B + R\, B\,R\, G) \nonumber \\ 
 && \,\, \& \, (R\, R\, B\, G - R\, R\, G\, B),
 \eea
with the left red petal as a starting point and dropping out all repetitions. 

In diagrams according to the Cayley–Hamilton theorem for a single $3 \times 3$ block one has 
{\ytableausetup {boxsize=1.7em}\begin{eqnarray}
J_{{a}^3}=\Tr(A^3) := 
\hspace{-0.7cm}{\raisebox{-1.6em}{  \begin{tikzpicture}
\hspace{0.2cm}{\raisebox{0.25em}{  \begin{polaraxis}[width=4.5cm,grid=none,  axis lines=none]
     	\addplot+[mark=none,domain=0:90,samples=50,color=black,style={ultra thick}] 
		{ sin(2*x)};
   \end{polaraxis}}
\hspace{0.65cm}{\raisebox{1.8em}{  \begin{polaraxis}[grid=none,  axis lines=none]
     	\addplot+[mark=none,domain=0:90,samples=50,color=black,style={ultra thick}] 
		{ sin(2*x)};
   \end{polaraxis}}}}
\hspace{-0.4cm}{\raisebox{0.9em}{ \begin{polaraxis}[width=4cm,grid=none,  axis lines=none]
     	\addplot+[mark=none,domain=0:90,samples=50,color=black,style={ultra thick}] 
		{ sin(2*x)};
   \end{polaraxis}}}
 \end{tikzpicture}}}\equiv \hspace{0.1cm}{\raisebox{-0.3cm}{  \begin{tikzpicture}
\begin{polaraxis}[grid=none,  axis lines=none]
     	\addplot+[mark=none,domain=0:270,samples=50,color=black,style={ultra thick}] 
		{sin(-3*x)}; 
   \end{polaraxis}
 \end{tikzpicture}}}\,.
\end{eqnarray}}
This is since producing higher order invariants terminated at the point where they can be rewritten in terms of the lower invariants. Otherwise, in the presence of another $3 \times 3$ matrix equivalent to another petal (that can be distinguished with a different sign or colour), always all petals of one type are adjacent so the invariant can be reduced i.e. 
\begin{align}
&J_{{a}^2{ab}^1}=\Tr(A^3B)={1\over 6} \Big[ \Tr(A^3)\Tr(B)
\nonumber \\&
-3 \Tr(A^2)\Tr(A) \Tr(B)+6 \Tr(A) \Tr(A^2B)
\nonumber \\&
+2\Tr(A)^3 \Tr(B)+3 \Tr(A B) \big(\Tr(A^2)-\Tr(A)^2\big)\Big],
\label{eq:A3C}
\end{align}
indicating the next invariant out of two building blocks 
\ytableausetup {boxsize=1.5em}\bea
J_{{a}^1{ab}^1}=\Tr(A^2B) := 
\hspace{-0.1cm}{\raisebox{-1.1em}{ 
\begin{tikzpicture}
 \begin{polaraxis}[grid=none,  axis lines=none]
     	\addplot+[mark=none,domain=0:60,samples=50,color=black,style={ultra thick},style=dashed] 
		{-sin(3*x)}; 
   \end{polaraxis}
   \begin{polaraxis}[grid=none,  axis lines=none]
     	\addplot+[mark=none,domain=61:180,samples=50,color=black,style={ultra thick}] 
		{-sin(3*x)}; 
		\end{polaraxis}
 \end{tikzpicture}}}
\eea
and similarly for $J_{{a b}^1{b}^1}=\Tr(A B^2)$. Additionally, out of two matrices, there is one non-reducible CP-conserving invariant that arises from
\begin{align}
J_{{a b}^2}&= \Tr ((AB)^2)+\Tr(A^2 B^2)=f(\rm Basic~Invs),
\end{align}
where equivalently this may be given as
\ytableausetup {boxsize=1.5em}\bea
J_{{ab}^2}=\Tr((A B)^2) := 
\hspace{-0.6cm}{\raisebox{-3.2em}{ 
\begin{tikzpicture}
\hspace{0.7cm}{\raisebox{1.8em}{  \begin{polaraxis}[grid=none,  axis lines=none]
     	\addplot+[mark=none,domain=0:90,samples=50,color=black,style={ultra thick}] 
		{ sin(2*x)};
   \end{polaraxis}}}
   \hspace{-0.3cm}{\raisebox{1.1em}{  \begin{polaraxis}[width=3.9cm,grid=none,  axis lines=none]
     	\addplot+[mark=none,domain=0:90,samples=50,color=black,style={ultra thick}] 
		{ sin(2*x)};
   \end{polaraxis}}}
   \hspace{-0.42cm}{\raisebox{0em}{ 
\hspace{0.7cm}{\raisebox{1.8em}{  \begin{polaraxis}[grid=none,  axis lines=none]
     	\addplot+[mark=none,domain=0:90,samples=50,color=black,style={ultra thick},style=dashed] 
		{ -sin(2*x)};
   \end{polaraxis}}}
   \hspace{-0.3cm}{\raisebox{1.1em}{  \begin{polaraxis}[width=3.9cm,grid=none,  axis lines=none]
     	\addplot+[mark=none,domain=0:90,samples=50,color=black,style={ultra thick},style=dashed] 
		{-sin(2*x)};
   \end{polaraxis}}}}}
 \end{tikzpicture}}}~~\equiv  \hspace{-0.1cm}{\raisebox{-1.3em}{ \begin{tikzpicture}
\begin{polaraxis}[grid=none, axis lines=none]
\addplot+[mark=none,domain=0:270,samples=50,color=black,style={ultra thick},style=dashed] 
		{-sin(2*x)};  
   \end{polaraxis}
 \begin{polaraxis}[grid=none, axis lines=none]
\addplot+[mark=none,domain=0:270,samples=50,color=black,style={ultra thick},style=dashed] 
		{+sin(2*x)};  
   \end{polaraxis}
      \begin{polaraxis}[grid=none, axis lines=none]
\addplot+[mark=none,domain=0:180,samples=50,color=black,style={ultra thick}] 
		{sin(2*x)};  
   \end{polaraxis}
\end{tikzpicture}}}\,.
\eea 
The highest invariant consisting of two matrices is the so-called ''Joint'' invariant. Here, the word ``Joint'' represents those invariants in which the summation between different permutations produces basic or lower-order joint invariants. 
In the case of two matrices, one may have the maximum petal set as
 $$\hspace{0.1cm}{\raisebox{-1.5em}{\begin{tikzpicture}
  \begin{polaraxis}[grid=none,  axis lines=none]
     	\addplot+[mark=none,domain=0:180,samples=50,color=black,style={ultra thick}] 
		{-sin(3*x)}; 
   \end{polaraxis}
   \begin{polaraxis}[grid=none,  axis lines=none]
     	\addplot+[mark=none,domain=0:180,samples=50,color=black,style={ultra thick},style=dashed] 
		{+sin(3*x)}; 
		\end{polaraxis}
 \end{tikzpicture}}}$$
pertinent to CP-even and CP-odd invariants. Although the CP-even invariant is reducible, as can be simply checked by rearranging petals or alternatively using the Cayley–Hamilton theorem, so
\begin{align}
& {J^+}_{{a b}^3}:=  \Tr ((A B)^3) \,+\,{3\over 2}~ \big[\Tr(A^2 B^2 A B)+\Tr(B^2 A^2 B A) \big]
\nonumber\\ 
& \equiv \; {1\over 2}\; \big[\Tr(A^2 B^2 A B)+\Tr(B^2 A^2 B A) \big]+f(\rm Basic~Invs)
\nonumber\\ 
& \equiv f(\rm Basic~Invs),
\end{align}  
and
 \bea
{J^-}_{{a b}^3}:=   \big[\Tr(A^2 B^2 A B)\,-\,\Tr(B^2 A^2 B A)\big],
 \label{eq:CP1}
\eea
where non-zero imaginary values of the above quantity indicate a non-vanishing CP-violating phase. 
Thus one CP-odd invariant in this order is considered. 

Furthermore, this procedure for the construction of higher-order invariants can be extended by including additional $3\times 3$ matrices. In this context, the lowest invariants involving an additional $3\times 3$ matrix $C$ become
 \bea
{J^\pm}_{a^1{b c}^1}:= \big[\Tr(A B C)\,\pm\,\Tr(A C B)\big]
\equiv   
\hspace{-0.1cm}{\raisebox{-1em}{\begin{tikzpicture}
  \begin{polaraxis}[grid=none,  axis lines=none]
     	\addplot+[mark=none,domain=61:120,samples=50,color=black,style={ultra thick}] 
		{-sin(3*x)}; 
   \end{polaraxis}
   \begin{polaraxis}[grid=none,  axis lines=none]
     	\addplot+[mark=none,domain=0:60,samples=50,color=black,style={ultra thick},style=dashed] 
		{-sin(3*x)}; 
		\end{polaraxis}
 \begin{polaraxis}[grid=none,  axis lines=none]
     	\addplot+[mark=none,domain=120:180,samples=50,color=black,style={ultra thick},style=dotted] 
		{-sin(3*x)}; 
\end{polaraxis}
 \end{tikzpicture}}}
 \label{eq:ABC}
 \eea
where $J^+$ and $J^-$ indicate CP-even and CP-odd invariants, respectively. Evidently, non-zero values of the CP-odd invariant introduce a non-vanishing CP-violating phase. Note that these relations are not reducible and can be regarded as basic invariants, however, for $2\times 2$ matrices these become reducible.

The next invariants out of three matrices $ A, B, C$ are joint invariants of order four rising from the following petals set
\bea  
\raisebox{-1em}{\begin{tikzpicture}
  \begin{polaraxis}[grid=none,  axis lines=none]
     	\addplot+[mark=none,domain=90:270,samples=50,color=black,style={ultra thick}] 
		{-sin(2*x)}; 
   \end{polaraxis}
   \begin{polaraxis}[grid=none,  axis lines=none]
     	\addplot+[mark=none,domain=0:90,samples=50,color=black,style={ultra thick},style=dashed] 
		{-sin(2*x)}; 
		\end{polaraxis}
 \begin{polaraxis}[grid=none,  axis lines=none]
     	\addplot+[mark=none,domain=270:360,samples=50,color=black,style={ultra thick},style=dotted] 
		{-sin(2*x)};
\end{polaraxis}
 \end{tikzpicture}}
 \nonumber
 \eea
 that is equivalent to two discrete forms
 \begin{align}
J_{{ab}^1{ac}^1}&+{J^+}_{a^2{bc}^1}=\Tr(A^2 B C)+\Tr(A^2 C B) +\Tr(A B A C)\nonumber\\
&=f(\rm Basic~Invs),
\label{ABAC}
 \end{align}
and a non-reducible CP-odd invariant
\begin{align}
{J^-}_{a^2{bc}^1}=\Tr(A^2 B C)\,-\,\Tr(A^2 C B).
\label{eq:AABC}
\end{align}
Therefore, based on this petal set there are a CP-odd and a CP-even invariants.

In a similar fashion, one can obtain higher-order combinations of three matrices and their permutations. For example, at order five for three blocks in a ring-diagram, one may have the following petals set
$$\hspace{0.2cm}{\raisebox{-1.4em}{\begin{tikzpicture}
  \begin{polaraxis}[grid=none,  axis lines=none]
     	\addplot+[mark=none,domain=216:324,samples=50,color=black,style={ultra thick}] 
		{-sin(5*x)}; 
   \end{polaraxis}
   \begin{polaraxis}[grid=none,  axis lines=none]
     	\addplot+[mark=none,domain=0:72,samples=50,color=black,style={ultra thick},style=dotted] 
		{-sin(5*x)}; 
		\end{polaraxis}
 \begin{polaraxis}[grid=none,  axis lines=none]
     	\addplot+[mark=none,domain=270:360,samples=50,color=black,style={ultra thick},style=dashed] 
		{-sin(5*x)}; 
\end{polaraxis}
 \end{tikzpicture}}}$$
 which can account only for one CP-odd invariant. This is because, for the CP-even case, three petals can always be placed adjacently
\begin{align}
&{J^+}_{a^3{cb}^1}+{J^+}_{a^1{ac}^1{ab}^1}= \Tr(A^3 C B)+\Tr(A^3 B C) 
\nonumber \\
& + 2 \Tr(A^2 B A C)+2 \Tr(A^2 C A B) 
 \nonumber \\
\sim& \Tr(A) \times f({\rm Joint\;Invs}) +f({\rm Basic\;Invs}) ,
\label{eq:A3BC}
\end{align}
and
\bea
{J^-}_{a^3{bc}^1}:=  \big[\Tr(A^2 B A C)\,-\,\Tr(A^2 C A B)\big].
\label{eq:AAABC}
 \eea
Thus one can account for only one CP-odd invariant in this type. Note that the CP-odd ${\rm Tr}(A^3 C B)-{\rm Tr}(A^3 B C)$ invariants can be written in terms of lower CP-odd invariants (i.e. the basic invariants ${\rm Tr}(A C B) - {\rm Tr}(A B C)$ and the CP-odd ${\rm Tr}(A^2 C B)- {\rm Tr}(A^2 B C)$). 

Moreover, in the presence of an additional $3\times 3$ matrix $D$, there are six permutations around $ABCD$ where their summation is reduced to basic invariants as
\begin{align}
\raisebox{-1.3em}{ \begin{tikzpicture}
   \begin{polaraxis}[grid=none, axis lines=none]
\addplot+[mark=none,domain=0:90,samples=50,color=black,style={ultra thick}] {sin(2*x)};  
   \end{polaraxis}
    \begin{polaraxis}[grid=none, axis lines=none]
\addplot+[mark=none,domain=90:180,samples=50,color=black,style={ultra thick},style=densely dashed] {sin(2*x)};  
   \end{polaraxis}
    \begin{polaraxis}[grid=none, axis lines=none]
\addplot+[mark=none,domain=180:270,samples=50,color=black,style={ultra thick},style=densely dashdotdotted] {sin(2*x)};  
   \end{polaraxis}
    \begin{polaraxis}[grid=none, axis lines=none]
\addplot+[mark=none,domain=270:360,samples=50,color=black,style={ultra thick},style=loosely dashed] {sin(2*x)};  
   \end{polaraxis}
 \end{tikzpicture}}&\,=\, {J_{{ab}^1{cd}^1}} + {J_{{bc}^1{ad}^1}}+ {J_{{ca }^1{bd}^1}}
\nonumber \\ &
= \Tr(A B C D)+ \Tr(A B D C) + \Tr(A C D B) \nonumber \\ 
& + \Tr(A C B D) 
 + \Tr(A D C B) + \Tr(A D B C)
\nonumber \\  &
=\,f(\rm Basic~Invs).
\label{ABCD}
\end{align}
Thus five of these permutations can be considered CP-even joint invariants. Note that the CP-odd relation is reducible to 
CP-odd invariants in the form of Eq.~\eqref{eq:ABC}.

This chain procedure is trivially extended up to higher orders, where only basic invariants and the summation of the joint invariant are present, as well as the summation of the lowest joint invariant that can be expressed in terms of basic invariants.

Up to four blocks from  Eqs.\eqref{eq:CP1}, \eqref{eq:ABC}, \eqref{eq:AABC} and \eqref{eq:AAABC}, one can realise in addition to the obvious CPV invariants via a complex matrix ${\rm Im}\Tr(A-{\rm h.c.})$ there are CP-odd invariants as
\begin{align}
    J^-_{{a b}^3} &= \text{Im} \Tr \big[ \{A^2,\, B^2\}[A,\, B] \big], \\
    J^-_{a^x b^y c^y} &= \text{Im} \Tr \big[ A^{x_0} [ A^{x_1} B^{y}, A^{x_1} C^{y} ] \big],
    \label{eq:CPs}
\end{align}
where $x=x_0+x_1$ stopped at power three. This approach can systematically be extended up to any $n$ order smaller than the termination point. Here, for the first time, we formulate the termination points of invariants based on the LO (Lowest Order) CP-odd due to matrix order (not their complex nature), HO (Highest Order) basic and LO joint invariants as shown in Table~\ref{tab:termination}.
\begin{table}[H]
\centering
\begin{tabular}{|l c c c |c|}
\hline
& LO CP-odd & HO Basic & LO Joint & Termination \\ \hline
& $n$ & $m$ & $k\geq m$ & $\geq n+k$ \\
& $n$ & $m$ & $k<m$ & $> n+m$ \\ \hline
\end{tabular}
\caption{The termination points of invariants.}
\label{tab:termination}
\end{table}
This way of formulation enables us to navigate CPV, thereby opening new pathways for future searches.

In the following section, we showcase the application of the ring-diagram in the framework of $\nu$SMEFT with operators of dimensions 5 and 6.
\section{Example: Invariants in the framework of $\nu$SMEFT with operators 5 and 6} \label{sec:SSMEFT}
In this section, we consider the framework of $\nu$SMEFT that includes three right-handed neutrinos $N_{\sf R}$, which are singlets under the SM gauge group~\cite{Minkowski:1977sc,Yanagida:1979as}.

The SM Lagrangian, $\mathcal{L_\mathrm{SM}}$, is extended through a series of higher-order dimension $n$ (dim-n) ${\rm SU(3)}_c\otimes{\rm SU(2)}_{\sf L}\otimes{\rm U(1)}_Y$ gauge-invariant operators, ${\rm O}_i^{({\rm dim}=n)}$, describing higher-order interactions as follows
 \begin{equation}
     \mathcal{L}_{\rm SMEFT}=\mathcal{L}_{\rm SM}+\sum_{n>4}\sum_i\frac{C_i^{({\rm dim}=n)}}{\Lambda^{n-4}}{\rm O}_i^{({\rm dim}=n)},
 \end{equation}
where $\Lambda$ is some higher mass scale and $C_i$ stands for the corresponding dimensionless coupling constants, i.e.~the so-called Wilson coefficients~\cite{eftlectures}. 

For an example, we take the $\nu$SMEFT that is extended to the order of dim-6. The leading flavor symmetry-breaking Lagrangian can be given by
\begin{eqnarray}
	\label{eq:EFT-lag6}
\mathcal{L}= {\mathcal L}_{\rm SM} -\left(\frac{C_5 }{2\Lambda} {\cal O}_5 + {\rm h.c.}\right)+\frac{C_6 }{\Lambda^2}{\cal O}_6 \;,
\end{eqnarray}
with the following dim-5 and dim-6 operators
\begin{eqnarray}
{\cal O}_5 =\overline L_{\sf L}\widetilde{H}\widetilde{H}^{\sf T}L_{\sf L}^{\sf C}\;,~
{\cal O}_6 =\left(\widetilde{H}^\dag i\overleftrightarrow{\partial}_\mu \widetilde{H}\right)\left(\overline L_i \gamma^\mu L_j\right),	
\end{eqnarray}
where ${L}_{\sf L} \equiv (\nu_{\sf L},\; l_{\sf L})^{\sf T}$ and $\mathcal{C}\equiv {\rm i}\gamma^2\gamma^0$ being the charge-conjugation operator. 
Accordingly, the corresponding Wilson coefficients read
\begin{eqnarray}
	\label{eq:Neutino-wilson}
	C_5 =-Y_\nu {\Lambda\over M_{\sf R}} Y_\nu^{\sf T}\;, \quad
	C_6 =Y_\nu   {\Lambda^2\over M_{\sf R}^\dagger M_{\sf R}}   Y_\nu^\dagger\;.
\end{eqnarray}
The transformation of Yukawa matrices, the Majorana masses ($M_{\sf R}$) and the Wilson coefficient is shown in Table~\ref{tab:tabdim56}.
\begin{table}[H]
\begin{center}
	\begin{tabular}{l|c|c|c}
		& \; $U(3)_{L}$ \; & \; $U(3)_e$ \; &\;  $U(3)_{\sf R}$ \; \\\hline
		$Y_e$ &$\mathbf{3}$ & $\mathbf{\overline{3}}$ & $\mathbf{1}$  \\[0.1cm]
		$Y_\nu$ &$\mathbf{3}$ & $ \mathbf{1}$ & $\mathbf{\overline{3}}$  \\[0.1cm]
		$ M_{\sf R}$ &$\mathbf{1}$ & $\mathbf{1}$ & $\mathbf{3}^*\times\mathbf{\overline{3}}$  \\[0.1cm]
		$C_5$ &$\mathbf{3}\times\mathbf{3}^{\sf T}$ & $\mathbf{1}$ & $\mathbf{1}$  \\
		$C_6$ &$\mathbf{3}\times\mathbf{3}^{\dagger}$ & $\mathbf{1}$ & $\mathbf{1}$  \\[0.1cm]
	\end{tabular}
	\end{center}
\caption{\it Flavor transformation under $U(3)_{L}$, $U(3)_e$ and $U(3)_{\sf R}$ of the Yukawa matrices $Y_{e,\nu}$, the Majorana masses $M_{\sf R}$ and the Wilson coefficients $C_5$ and $C_{6}$.}
	\label{tab:tabdim56}
\end{table}

\begin{figure}[H]
\begin{center}
\includegraphics[width=0.3\textwidth]{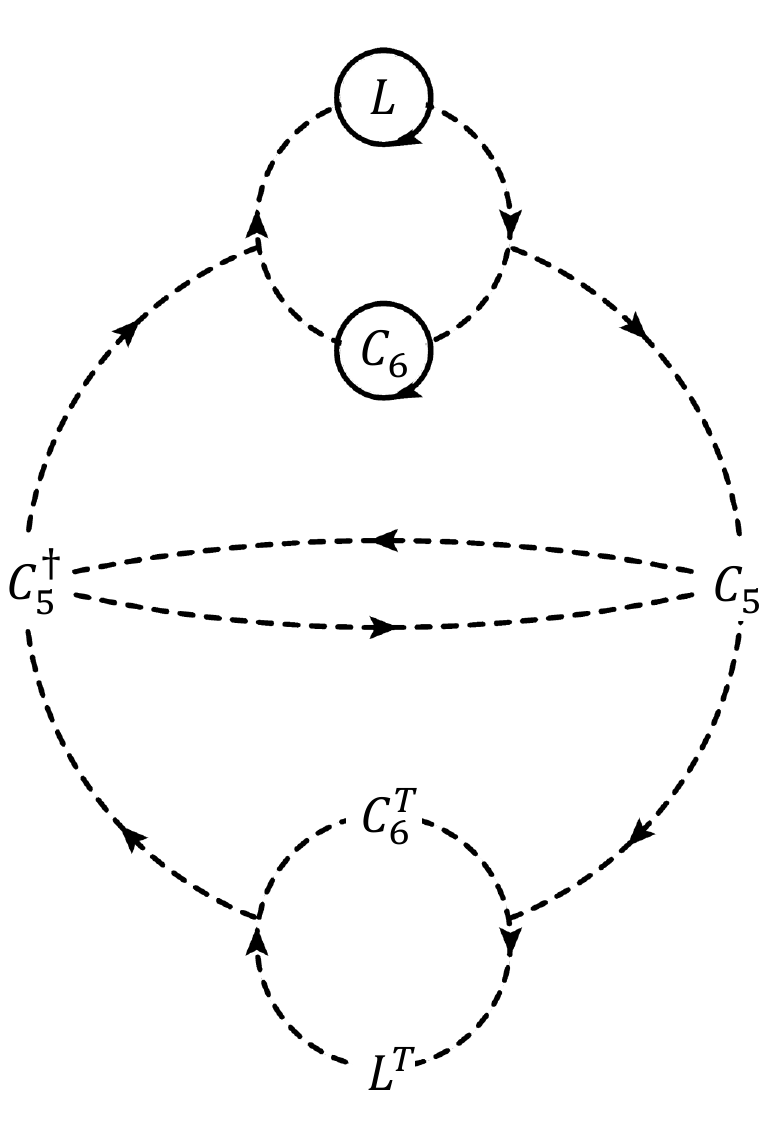}
\end{center}
\caption{The Ring-diagram for $\nu$SMEFT with Wilson coefficients $C_{5,6}$.}
\label{RD7}
\end{figure}

Skipping the first round of trivial blocks out of the ring-diagram, the building blocks for the construction of flavour invariants are $\left\{L \equiv Y_eY_e^\dagger, C_5, C_6 \right\}$ as can be seen in Ring-diagram~\ref{RD7}.

Consequently, the building blocks are
\begin{subequations}
{\allowdisplaybreaks
\begin{eqnarray}
 M_l &\equiv & \text{diag} \begin{pmatrix} L & 0  \end{pmatrix}\equiv  \hspace{0.2cm}{\raisebox{-0.7em}{\begin{tikzpicture}
\def \radius {0.4cm}
  \node[draw, circle,style={thick},-latex] {$L$};
  \draw[-latex,->,thick] ({180}:\radius);
\end{tikzpicture}}}\,,
\\
M_{c_6} &\equiv & \text{diag} \begin{pmatrix} 0 & C_6  \end{pmatrix}\equiv  \hspace{0.2cm}{\raisebox{-0.7em}{\begin{tikzpicture}
\def \radius {0.4cm}
  \node[draw, circle,style={thick},-latex] {$\scriptstyle{C_6}$};
  \draw[-latex,->,thick] ({180}:\radius);
\end{tikzpicture}}}\,,
\\
 {M_{c}} &\equiv &  \begin{pmatrix}
C_5\,C_5 ^{\dagger} & 0 \\
0 & \, C_5 ^{\dagger}\,C_5   \\
\end{pmatrix}~ \equiv \raisebox{-1em}{ 
\begin{tikzpicture}[transform shape,line width=0.9pt]
\node (A) at (0,0) {$C_5$}; 
\node (B) at (1.2,0) {$C_5^{\dagger}$}; 
\path[draw] 
(A) edge[bend right=-40,black,dashed,->] (B)
(B) edge[bend right=-40,black,-latex,dashed,->] (A);
\end{tikzpicture}}
\nonumber\\ &\equiv & 
\hspace{0.2cm}{\raisebox{-0.7em}{\begin{tikzpicture}
\def \radius {0.4cm}
  \node[draw, circle,style={thick},-latex] {$C$};
  \draw[-latex,->,thick] ({180}:\radius);
\end{tikzpicture}}}\,,\\
 {M_{lc_6}}^\pm &\equiv &   \begin{pmatrix}
L\,C_6 & 0 \\
0 & \,\pm C_6 L   \\
\end{pmatrix}~ \equiv \raisebox{-1em}{ 
\begin{tikzpicture}[transform shape,line width=0.9pt]
\node (A) at (0,0) {$L$}; 
\node (B) at (1.2,0) { $C_6$ }; 
\path[draw] 
(A) edge[bend right=-40,black,dashed,->] (B)
(B) edge[bend right=-40,black,-latex,dashed,->] (A);
\end{tikzpicture}}\,,
 \\
M_{lk_m}^\pm &\equiv &  \begin{pmatrix}
L \,K_m& 0\\
0 &\,  \pm K_m\,L  \\
\end{pmatrix}
\equiv  \raisebox{-1em}{ 
\begin{tikzpicture}[transform shape,line width=0.9pt]
\node (A) at (0,0) {$L$}; 
\node (B) at (1.2,0) { $K_m$ }; 
\path[draw] 
(A) edge[bend right=-40,black,dashed,->] (B)
(B) edge[bend right=-40,black,-latex,dashed,->] (A);
\end{tikzpicture}} \nonumber \\  &\equiv &  \raisebox{-0.1em}{ 
\raisebox{-2.8em}{ \begin{tikzpicture}[transform shape,line width=1pt]
\node (A) at (0.9,1) {$C_5^{\dagger}$}; 
\node (B) at (2,2) {$L$}; 
\node (C) at (3,1) {$C_5$};
\node (D) at (2,0) {${L^{\sf T}}^m$}; 
\node (E) at (1,0.8) {}; 
\path[draw] 
(A) edge[bend right=-30,black,dashed,->] (B) 
(B) edge[bend right=-30,black,dashed,->] (C)
(C) edge[bend right=-30,black,dashed,->] (D) 
(D) edge[bend right=-30,black,dashed,->] (E);
\end{tikzpicture}}}\,,
\\
M_{c_6k_m}^\pm &\equiv &   \begin{pmatrix}
C_6\,K_m& 0\\
0 &\,  \pm K_m\,C_6 \\
\end{pmatrix}~ \equiv  \raisebox{-1em}{ 
\begin{tikzpicture}[transform shape,line width=0.9pt]
\node (A) at (0,0) {$C_6$}; 
\node (B) at (1.2,0) { $K_m$ }; 
\path[draw] 
(A) edge[bend right=-40,black,dashed,->] (B)
(B) edge[bend right=-40,black,-latex,dashed,->] (A);
\end{tikzpicture}} \nonumber\\
\,&\equiv& \raisebox{-0.1em}{ 
\raisebox{-2.8em}{ \begin{tikzpicture}[transform shape,line width=1pt]
\node (A) at (0.9,1) {$C_5^{\dagger}$}; 
\node (B) at (2,2) {$C_6$}; 
\node (C) at (3,1) {$C_5$};
\node (D) at (2,0) {${L^{\sf T}}^m$}; 
\node (E) at (1,0.8) {}; 
\path[draw] 
(A) edge[bend right=-30,black,dashed,->] (B) 
(B) edge[bend right=-30,black,dashed,->] (C)
(C) edge[bend right=-30,black,dashed,->] (D) 
(D) edge[bend right=-30,black,dashed,->] (E);
\end{tikzpicture}}}\,,
\\
M_{lh_m}^\pm &\equiv &   \begin{pmatrix}
L \,H_m& 0\\
0 &\,  \pm H_m\,L  \\
\end{pmatrix}~ \equiv  \raisebox{-1em}{ 
\begin{tikzpicture}[transform shape,line width=0.9pt]
\node (A) at (0,0) {$L$}; 
\node (B) at (1.2,0) { $H_m$ }; 
\path[draw] 
(A) edge[bend right=-40,black,dashed,->] (B)
(B) edge[bend right=-40,black,-latex,dashed,->] (A);
\end{tikzpicture}} \nonumber \\\,&\equiv&\raisebox{-0.1em}{ 
\raisebox{-2.8em}{ \begin{tikzpicture}[transform shape,line width=1pt]
\node (A) at (0.9,1) {$C_5^{\dagger}$}; 
\node (B) at (2,2) {$L$}; 
\node (C) at (3,1) {$C_5$};
\node (D) at (2,0) {${{C_6}^{\sf T}}^m$}; 
\node (E) at (1,0.8) {}; 
\path[draw] 
(A) edge[bend right=-30,black,dashed,->] (B) 
(B) edge[bend right=-30,black,dashed,->] (C)
(C) edge[bend right=-30,black,dashed,->] (D) 
(D) edge[bend right=-30,black,dashed,->] (E);
\end{tikzpicture}}}\,,
\\
M_{c_6H_m}^\pm &\equiv &   \begin{pmatrix}
C_6 \,H_m& 0\\
0 &\,  \pm H_m\,C_6  \\
\end{pmatrix}~ \equiv  \raisebox{-1em}{ 
\begin{tikzpicture}[transform shape,line width=0.9pt]
\node (A) at (0,0) {$C_6$}; 
\node (B) at (1.2,0) { $H_m$ }; 
\path[draw] 
(A) edge[bend right=-40,black,dashed,->] (B)
(B) edge[bend right=-40,black,-latex,dashed,->] (A);
\end{tikzpicture}}
\nonumber \\
\,&\equiv& \raisebox{-0.1em}{ 
\raisebox{-2.8em}{ \begin{tikzpicture}[transform shape,line width=1pt]
\node (A) at (0.9,1) {$C_5^{\dagger}$}; 
\node (B) at (2,2) {$C_6$}; 
\node (C) at (3,1) {$C_5$};
\node (D) at (2,0) {${{C_6}^{\sf T}}^m$}; 
\node (E) at (1,0.8) {}; 
\path[draw] 
(A) edge[bend right=-30,black,dashed,->] (B) 
(B) edge[bend right=-30,black,dashed,->] (C)
(C) edge[bend right=-30,black,dashed,->] (D) 
(D) edge[bend right=-30,black,dashed,->] (E);
\end{tikzpicture}}}\,,
 \\
M_{l{f_{mn}}^\pm}^\pm &\equiv &   \begin{pmatrix}
L \,{F_{mn}}^\pm& 0\\
0 &\,  \pm {F_{mn}}^\pm\,L  \\
\end{pmatrix}~ \equiv  \raisebox{-1em}{ 
\begin{tikzpicture}[transform shape,line width=0.9pt]
\node (A) at (0,0) {$L$}; 
\node (B) at (1.2,0) { ${F_{mn}}^\pm$ }; 
\path[draw] 
(A) edge[bend right=-40,black,dashed,->] (B)
(B) edge[bend right=-40,black,-latex,dashed,->] (A);
\end{tikzpicture}} \nonumber \\ \,&\equiv& \raisebox{-0.1em}{ 
\raisebox{-2.8em}{ \begin{tikzpicture}[transform shape,line width=1pt]
\node (A) at (0.9,1) {$C_5^{\dagger}$}; 
\node (B) at (2.2,2.2) {$L$}; 
\node (C) at (3.3,1) {$C_5$};
\node (D) at (2.2,0) {${{X}^{mn}}^\pm$}; 
\node (E) at (1,0.7) {}; 
\path[draw] 
(A) edge[bend right=-30,black,dashed,->] (B) 
(B) edge[bend right=-30,black,dashed,->] (C)
(C) edge[bend right=-30,black,dashed,->] (D) 
(D) edge[bend right=-30,black,dashed,->] (E);
\end{tikzpicture}}}\,,
 \label{blocks-dim-6-2}
\end{eqnarray}}
\end{subequations}
where  $K_m=C_5 \,{L^{\sf T}}^m\,C_5^{\dagger}$,  $H_m=C_5 \,{C_6^{\sf T}}^m\,C_5^{\dagger}$ and ${F_{mn}}^\pm=C_5 \,{X^{mn}}^\pm\,C_5^{\dagger}$ with ${X^{mn}}^\pm={L^{\sf T}}^m {C_6^{\sf T}}^n \pm{C_6^{\sf T}}^n{L^{\sf T}}^m$ running over $m,n=1,2$. Additionally, the blocks $M_{c_6H_m}^\pm$ out of ring $ \raisebox{-1em}{ 
\begin{tikzpicture}[transform shape,line width=0.9pt]
\node (A) at (0,0) {$C_6$}; 
\node (B) at (1.2,0) { $H_m$ }; 
\path[draw] 
(A) edge[bend right=-40,black,dashed,->] (B)
(B) edge[bend right=-40,black,-latex,dashed,->] (A);
\end{tikzpicture}}$ can be skipped as it generates dim-10 invariants; however, we retain this block for comparative analysis with the number deduced through the Hilbert series approach.

One can organise the basic invariants beginning with the lowest order blocks as follows:

$\bullet$ Order-1:
{\allowdisplaybreaks \ytableausetup{smalltableaux}
\bea
J^+_{l^1}:=&  \hspace{0.0em}{ \begin{tikzpicture}
   \begin{polaraxis}[grid=none, axis lines=none]
     	\addplot+[mark=none,domain=0:90,  samples=15,color=pink,style={ultra thick}] 
		{sin(2*x)}; 
   \end{polaraxis}
 \end{tikzpicture}}=  \Tr(L), \quad  
 J^+_{c_6^1}:=&  \hspace{0.0em}{ \begin{tikzpicture}
   \begin{polaraxis}[grid=none, axis lines=none]
     	\addplot+[mark=none,domain=0:90,  samples=15,color=lime,style={ultra thick}] 
		{sin(2*x)}; 
   \end{polaraxis}
 \end{tikzpicture}}= \Tr(C_6).
 \nonumber 
\eea }

$\bullet$ Order-2:
{\allowdisplaybreaks \ytableausetup{smalltableaux}
\bea
J^+_{l^2}:=& \hspace{0.0em}{\raisebox{-1.5
em}{\begin{tikzpicture}
   \begin{polaraxis}[grid=none, axis lines=none]
     	\addplot+[mark=none,domain=0:90,  samples=15,color=pink,style={ultra thick}] 
		{sin(2*x)}; 
   \end{polaraxis}
    \begin{polaraxis}[grid=none, axis lines=none]
     	\addplot+[mark=none,domain=0:90,  samples=15,color=pink,style={ultra thick}] 
		{-sin(2*x)}; 
   \end{polaraxis}
 \end{tikzpicture}}}
=\Tr({L}^2),
\quad
J^+_{c^1} :=& \hspace{0.0cm}{\raisebox{0.2em}{ \begin{tikzpicture}\begin{polaraxis}[width=3.6cm,grid=none,  axis lines=none]	\addplot+[mark=none,domain=0:90,  samples=15,color=darkgreen,style=dashed,style={ultra thick}] {sin(2*x)};\end{polaraxis} \end{tikzpicture}}}
= \Tr (|C_5|^2),
\nonumber \\
J^+_{C_6^2}:=& \hspace{0.0em}{\raisebox{-1.5
em}{\begin{tikzpicture}
   \begin{polaraxis}[grid=none, axis lines=none]
     	\addplot+[mark=none,domain=0:90,  samples=15,color=lime,style={ultra thick}] 
		{sin(2*x)}; 
   \end{polaraxis}
    \begin{polaraxis}[grid=none, axis lines=none]
     	\addplot+[mark=none,domain=0:90,  samples=15,color=lime,style={ultra thick}] 
		{-sin(2*x)}; 
   \end{polaraxis}
 \end{tikzpicture}}}
=\Tr({C_6}^2),
\quad
J^+_{lc_6} :=&  \hspace{-0.2cm}{\raisebox{-1.5em}{ \begin{tikzpicture}
   \begin{polaraxis}[grid=none, axis lines=none]
     	\addplot+[mark=none,domain=0:90,  samples=15,color=pink,style={ultra thick}] 
		{sin(2*x)}; 
   \end{polaraxis}
    \begin{polaraxis}[grid=none, axis lines=none]
     	\addplot+[mark=none,domain=0:90,  samples=15,color=lime,style={ultra thick}] 
		{-sin(2*x)}; 
   \end{polaraxis}
\end{tikzpicture}}}
= \Tr(LC_6).
\nonumber 
\eea }    

$\bullet$ Order-3:
{\allowdisplaybreaks \ytableausetup{smalltableaux}
\bea
J_{l^3} :=& \hspace{-0.1cm}{\raisebox{-1.1em}{\begin{tikzpicture}
 \begin{polaraxis}[grid=none,  axis lines=none]
     	\addplot+[mark=none,domain=0:270, samples=30,color=pink,style={ultra thick}] 
		{-sin(3*x)}; 
   \end{polaraxis}
 \end{tikzpicture}}}=\Tr ({L}^3),~~~~
\nonumber \\ 
J^+_{l^1c^1} :=&  \hspace{0.0em}{\raisebox{-1.5
em}{\begin{tikzpicture}
   \begin{polaraxis}[grid=none, axis lines=none]
     	\addplot+[mark=none,domain=0:90,  samples=15,color=pink,style={ultra thick}] 
		{sin(2*x)}; 
   \end{polaraxis}
    \begin{polaraxis}[grid=none, axis lines=none]
     	\addplot+[mark=none,domain=0:90,  samples=15,color=darkgreen,style={ultra thick},style=dashed] 
		{-sin(2*x)}; 
   \end{polaraxis}
 \end{tikzpicture}}} = \Tr(L|C_5|^2),~~
 \nonumber \\
J_{c_6^3} :=& \hspace{-0.1cm}{\raisebox{-1.1em}{\begin{tikzpicture}
 \begin{polaraxis}[grid=none,  axis lines=none]
     	\addplot+[mark=none,domain=0:270, samples=30,color=lime,style={ultra thick}] 
		{-sin(3*x)}; 
   \end{polaraxis}
 \end{tikzpicture}}}=\Tr ({C_6}^3)~,~~
 \nonumber \\
J^+_{lc_6^1{c_6}^1} :=&   \hspace{0.0em}{\raisebox{-1.5
em}{\begin{tikzpicture}
   \begin{polaraxis}[grid=none, axis lines=none]
     	\addplot+[mark=none,domain=0:60,  samples=15,color=lime,style={ultra thick}] 
		{-sin(3*x)}; 
   \end{polaraxis}
    \begin{polaraxis}[grid=none, axis lines=none]
     	\addplot+[mark=none,domain=60:120,  samples=15,color=lime,style={ultra thick}] 
		{-sin(3*x)}; 
   \end{polaraxis}
    \begin{polaraxis}[grid=none, axis lines=none]
     	\addplot+[mark=none,domain=120:180,  samples=15,color=pink,style={ultra thick}] 
		{-sin(3*x)}; 
   \end{polaraxis}
 \end{tikzpicture}}} = \Tr(L C_6^2),
 \nonumber \\
J_{l^1lc_6^1} :=& \hspace{-0.1cm}{\raisebox{-1.1em}{\begin{tikzpicture}
 \begin{polaraxis}[grid=none, axis lines=none]
     	\addplot+[mark=none,domain=0:60,  samples=15,color=pink,style={ultra thick}] 
		{-sin(3*x)}; 
   \end{polaraxis}
    \begin{polaraxis}[grid=none, axis lines=none]
     	\addplot+[mark=none,domain=60:120,  samples=15,color=pink,style={ultra thick}] 
		{-sin(3*x)}; 
   \end{polaraxis}
    \begin{polaraxis}[grid=none, axis lines=none]
     	\addplot+[mark=none,domain=120:180,  samples=15,color=lime,style={ultra thick}] 
		{-sin(3*x)}; 
   \end{polaraxis}
 \end{tikzpicture}}}=\Tr ({L}^2 C_6),~~
  \nonumber \\
J^+_{c_6^1c^1} :=&  \hspace{0.0em}{\raisebox{-1.5
em}{\begin{tikzpicture}
   \begin{polaraxis}[grid=none, axis lines=none]
     	\addplot+[mark=none,domain=0:90,  samples=15,color=lime,style={ultra thick}] 
		{sin(2*x)}; 
   \end{polaraxis}
    \begin{polaraxis}[grid=none, axis lines=none]
     	\addplot+[mark=none,domain=0:90,  samples=15,color=darkgreen,style={ultra thick},style=dashed] 
		{-sin(2*x)}; 
   \end{polaraxis}
 \end{tikzpicture}}} = \Tr(C_6|C_5|^2).~~~
  \label{order-3-dim5-2}
  \nonumber 
\eea }

$\bullet$ Order-4:
{\allowdisplaybreaks \ytableausetup{smalltableaux}
\bea
J^+_{l^2{c}^1} &:=& \hspace{-0.1cm}{\raisebox{-1.1em}{\begin{tikzpicture}
 \begin{polaraxis}[grid=none,  axis lines=none]
     	\addplot+[mark=none,domain=120:180,  samples=15,color=pink,style={ultra thick}] 
		{-sin(3*x)}; 
   \end{polaraxis}
    \begin{polaraxis}[grid=none,  axis lines=none]
     	\addplot+[mark=none,domain=60:120,  samples=15,color=pink,style={ultra thick}] 
		{-sin(3*x)}; 
   \end{polaraxis}
       \begin{polaraxis}[grid=none,  axis lines=none]
     	\addplot+[mark=none,domain=0:60,  samples=15,color=darkgreen,style={ultra thick},style=dashed] 
		{-sin(3*x)}; 
   \end{polaraxis}
 \end{tikzpicture}}} =\Tr(L^2|C_5|^2),
 \nonumber \\
J^+_{{c}^2}  &:=& \hspace{0.0em}{\raisebox{-1.5
em}{\begin{tikzpicture}
   \begin{polaraxis}[grid=none, axis lines=none]
     	\addplot+[mark=none,domain=0:90,  samples=15,color=darkgreen,style={ultra thick},style=dashed] 
		{sin(2*x)}; 
   \end{polaraxis}
    \begin{polaraxis}[grid=none, axis lines=none]
     	\addplot+[mark=none,domain=0:90,  samples=15,color=darkgreen,style={ultra thick},style=dashed] 
		{-sin(2*x)}; 
   \end{polaraxis}
 \end{tikzpicture}}}=\Tr(|C_5|^4),
\nonumber \\
J^+_{{lk_1}^1}&:=&  \hspace{-0.2cm}{\raisebox{-1.4em}{ \begin{tikzpicture}
     \begin{polaraxis}[grid=none, axis lines=none]
\addplot+[mark=none,domain=0:90,  samples=15,color=pink,style={ultra thick}] 
		{+sin(2*x)};  
   \end{polaraxis}
    \begin{polaraxis}[grid=none, axis lines=none]
\addplot+[mark=none,domain=0:90,  samples=15,color=orange,style={ultra thick},style=dashed] 
		{-sin(2*x)};  
   \end{polaraxis}
    \end{tikzpicture}}} = \Tr (L C_5 L^{\sf T} C_5^\dagger),
 \nonumber \\
 J^+_{c_6^2{c}^1}  &:=& \hspace{-0.1cm}{\raisebox{-1.1em}{\begin{tikzpicture}
 \begin{polaraxis}[grid=none,  axis lines=none]
     	\addplot+[mark=none,domain=120:180,  samples=15,color=lime,style={ultra thick}] 
		{-sin(3*x)}; 
   \end{polaraxis}
    \begin{polaraxis}[grid=none,  axis lines=none]
     	\addplot+[mark=none,domain=60:120,  samples=15,color=lime,style={ultra thick}] 
		{-sin(3*x)}; 
   \end{polaraxis}
       \begin{polaraxis}[grid=none,  axis lines=none]
     	\addplot+[mark=none,domain=0:60,  samples=15,color=darkgreen,style={ultra thick},style=dashed] 
		{-sin(3*x)}; 
   \end{polaraxis}
 \end{tikzpicture}}} =\Tr(C_6^2|C_5|^2),
 \nonumber \\
J^+_{lc_6^2} &:=&  \hspace{-0.1cm}{\raisebox{-1.1em}{\begin{tikzpicture}
 \begin{polaraxis}[grid=none,  axis lines=none]
     	\addplot+[mark=none,domain=0:180,  samples=15,color=pink,style={ultra thick}] 
		{sin(2*x)}; 
   \end{polaraxis}
    \begin{polaraxis}[grid=none,  axis lines=none]
     	\addplot+[mark=none,domain=180:360,  samples=15,color=lime,style={ultra thick}] 
		{sin(2*x)}; 
   \end{polaraxis}
 \end{tikzpicture}}} =\Tr(L^2 C_6^2),
 \nonumber \\
J^+_{{c_6k_1}^1} &:=&\hspace{-0.2cm}{\raisebox{-1.4em}{ \begin{tikzpicture}
     \begin{polaraxis}[grid=none, axis lines=none]
\addplot+[mark=none,domain=0:90,  samples=15,color=lime,style={ultra thick}] 
		{+sin(2*x)};  
   \end{polaraxis}
    \begin{polaraxis}[grid=none, axis lines=none]
\addplot+[mark=none,domain=0:90,  samples=15,color=orange,style={ultra thick},style=dashed] 
		{-sin(2*x)};  
   \end{polaraxis}
    \end{tikzpicture}}} = \Tr (C_6 C_5 L^{\sf T} C_5^\dagger),
\nonumber \\
J^+_{{c_6h_1}^1}&:=&  \hspace{-0.2cm}{\raisebox{-1.4em}{ \begin{tikzpicture}
     \begin{polaraxis}[grid=none, axis lines=none]
\addplot+[mark=none,domain=0:90,  samples=15,color=lime,style={ultra thick}] 
		{+sin(2*x)};  
   \end{polaraxis}
    \begin{polaraxis}[grid=none, axis lines=none]
\addplot+[mark=none,domain=0:90,  samples=15,color=olive,style={ultra thick},style=dashed] 
		{-sin(2*x)};  
   \end{polaraxis}
    \end{tikzpicture}}} = \Tr (C_6 C_5 C_6^{\sf T} C_5^\dagger),
 \nonumber \\
J^+_{{lc_6}^1c^1}  &:=&  \hspace{-0.1cm}{\raisebox{-1.1em}{\begin{tikzpicture}
 \begin{polaraxis}[grid=none,  axis lines=none]
     	\addplot+[mark=none,domain=120:180,  samples=15,color=lime,style={ultra thick}] 
		{-sin(3*x)}; 
   \end{polaraxis}
    \begin{polaraxis}[grid=none,  axis lines=none]
     	\addplot+[mark=none,domain=60:120,  samples=15,color=pink,style={ultra thick}] 
		{-sin(3*x)}; 
   \end{polaraxis}
       \begin{polaraxis}[grid=none,  axis lines=none]
     	\addplot+[mark=none,domain=0:60,  samples=15,color=darkgreen,style={ultra thick},style=dashed] 
		{-sin(3*x)}; 
   \end{polaraxis}
 \end{tikzpicture}}} =\Tr(L C_6 |C_5|^2 )\;+\; \Tr( L |C_5|^2 C_6 ), \qquad \qquad \qquad
 \nonumber
\eea}
where the dashed orange and olive petals presenting order three blocks $K_1 \equiv C_5 {L^{\sf T}} C_5^\dagger$ and $H_1 \equiv C_5 {C_6^{\sf T}} C_5^\dagger$, respectively.

$\bullet$ Order-5:
{\allowdisplaybreaks \ytableausetup{smalltableaux}
\bea
J^+_{l^1c^2} &:=&  \hspace{-0.1cm}{\raisebox{-1.1em}{\begin{tikzpicture}
 \begin{polaraxis}[grid=none,  axis lines=none]
     	\addplot+[mark=none,domain=60:120,  samples=15,color=pink,style={ultra thick}] 
		{-sin(3*x)}; 
   \end{polaraxis}
    \begin{polaraxis}[grid=none,  axis lines=none]
     	\addplot+[mark=none,domain=120:180,  samples=15,color=darkgreen,style={ultra thick},style=dashed] 
		{-sin(3*x)}; 
   \end{polaraxis}
       \begin{polaraxis}[grid=none,  axis lines=none]
     	\addplot+[mark=none,domain=0:60,  samples=15,color=darkgreen,style={ultra thick},style=dashed] 
		{-sin(3*x)}; 
   \end{polaraxis}
 \end{tikzpicture}}} =\Tr(L |C_5|^4), 
 \nonumber \\
J^+_{l^1{lk_1}^1} &:=&  \hspace{-0.1cm}{\raisebox{-1.1em}{\begin{tikzpicture}
 \begin{polaraxis}[grid=none,  axis lines=none]
     	\addplot+[mark=none,domain=60:120,  samples=15,color=pink,style={ultra thick}] 
		{-sin(3*x)}; 
   \end{polaraxis}
    \begin{polaraxis}[grid=none,  axis lines=none]
     	\addplot+[mark=none,domain=120:180,  samples=15,color=pink,style={ultra thick}] 
		{-sin(3*x)}; 
   \end{polaraxis}
       \begin{polaraxis}[grid=none,  axis lines=none]
     	\addplot+[mark=none,domain=0:60,  samples=15,color=orange,style={ultra thick},style=dashed] 
		{-sin(3*x)}; 
   \end{polaraxis}
 \end{tikzpicture}}}= \Tr (L^2 C_5 L^{\sf T} C_5^\dagger),
 \nonumber \\
 J^+_{l^1c^2} &:=& \hspace{-0.1cm}{\raisebox{-1.1em}{\begin{tikzpicture}
 \begin{polaraxis}[grid=none,  axis lines=none]
     	\addplot+[mark=none,domain=60:120,  samples=15,color=lime,style={ultra thick}] 
		{-sin(3*x)}; 
   \end{polaraxis}
    \begin{polaraxis}[grid=none,  axis lines=none]
     	\addplot+[mark=none,domain=120:180,  samples=15,color=darkgreen,style={ultra thick},style=dashed] 
		{-sin(3*x)}; 
   \end{polaraxis}
       \begin{polaraxis}[grid=none,  axis lines=none]
     	\addplot+[mark=none,domain=0:60,  samples=15,color=darkgreen,style={ultra thick},style=dashed] 
		{-sin(3*x)}; 
   \end{polaraxis}
 \end{tikzpicture}}} =\Tr(C_6 |C_5|^4), 
 \nonumber \\
 J^+_{l^1{lh_1}^1}  &:=& \hspace{-0.1cm}{\raisebox{-1.1em}{\begin{tikzpicture}
 \begin{polaraxis}[grid=none,  axis lines=none]
     	\addplot+[mark=none,domain=60:120,  samples=15,color=pink,style={ultra thick}] 
		{-sin(3*x)}; 
   \end{polaraxis}
    \begin{polaraxis}[grid=none,  axis lines=none]
     	\addplot+[mark=none,domain=120:180,  samples=15,color=pink,style={ultra thick}] 
		{-sin(3*x)}; 
   \end{polaraxis}
       \begin{polaraxis}[grid=none,  axis lines=none]
     	\addplot+[mark=none,domain=0:60,  samples=15,color=olive,style={ultra thick},style=dashed] 
		{-sin(3*x)}; 
   \end{polaraxis}
 \end{tikzpicture}}}= \Tr (L^2 C_5 C_6^{\sf T} C_5^\dagger),
 \nonumber \\
 J^+_{c_6^1{c_6k_1}^1} &:=& \hspace{-0.1cm}{\raisebox{-1.1em}{\begin{tikzpicture}
 \begin{polaraxis}[grid=none,  axis lines=none]
     	\addplot+[mark=none,domain=60:120,  samples=15,color=lime,style={ultra thick}] 
		{-sin(3*x)}; 
   \end{polaraxis}
    \begin{polaraxis}[grid=none,  axis lines=none]
     	\addplot+[mark=none,domain=120:180,  samples=15,color=lime,style={ultra thick}] 
		{-sin(3*x)}; 
   \end{polaraxis}
       \begin{polaraxis}[grid=none,  axis lines=none]
     	\addplot+[mark=none,domain=0:60,  samples=15,color=orange,style={ultra thick},style=dashed] 
		{-sin(3*x)}; 
   \end{polaraxis}
 \end{tikzpicture}}}= \Tr (C_6^2 C_5 L^{\sf T} C_5^\dagger),
 \nonumber \\
 J^+_{c_6^1{c_6h_1}^1}  &:=&  \hspace{-0.1cm}{\raisebox{-1.1em}{\begin{tikzpicture}
 \begin{polaraxis}[grid=none,  axis lines=none]
     	\addplot+[mark=none,domain=60:120,  samples=15,color=lime,style={ultra thick}] 
		{-sin(3*x)}; 
   \end{polaraxis}
    \begin{polaraxis}[grid=none,  axis lines=none]
     	\addplot+[mark=none,domain=120:180,  samples=15,color=lime,style={ultra thick}] 
		{-sin(3*x)}; 
   \end{polaraxis}
       \begin{polaraxis}[grid=none,  axis lines=none]
     	\addplot+[mark=none,domain=0:60,  samples=15,color=olive,style={ultra thick},style=dashed] 
		{-sin(3*x)}; 
   \end{polaraxis}
 \end{tikzpicture}}}= \Tr ({C_6}^2 C_5 C_6^{\sf T} C_5^\dagger),
 \nonumber \\
 J^+_{l^1{c_6k_1}^1} &:= &  \hspace{-0.1cm}{\raisebox{-1.1em}{\begin{tikzpicture}
 \begin{polaraxis}[grid=none,  axis lines=none]
     	\addplot+[mark=none,domain=60:120,  samples=15,color=pink,style={ultra thick}] 
		{-sin(3*x)}; 
   \end{polaraxis}
    \begin{polaraxis}[grid=none,  axis lines=none]
     	\addplot+[mark=none,domain=120:180,  samples=15,color=lime,style={ultra thick}] 
		{-sin(3*x)}; 
   \end{polaraxis}
       \begin{polaraxis}[grid=none,  axis lines=none]
     	\addplot+[mark=none,domain=0:60,  samples=15,color=orange,style={ultra thick},style=dashed] 
		{-sin(3*x)}; 
   \end{polaraxis}
 \end{tikzpicture}}}
 \nonumber \\ & = & \Tr (L C_6 C_5 L^{\sf T} C_5^\dagger)+\Tr (C_6 L C_5 L^{\sf T} C_5^\dagger),
 \nonumber \\
  J^+_{l^1{c_6h_1}^1}&:=&  \hspace{-0.1cm}{\raisebox{-1.1em}{\begin{tikzpicture}
 \begin{polaraxis}[grid=none,  axis lines=none]
     	\addplot+[mark=none,domain=60:120,  samples=15,color=pink,style={ultra thick}] 
		{-sin(3*x)}; 
   \end{polaraxis}
    \begin{polaraxis}[grid=none,  axis lines=none]
     	\addplot+[mark=none,domain=120:180,  samples=15,color=lime,style={ultra thick}] 
		{-sin(3*x)}; 
   \end{polaraxis}
       \begin{polaraxis}[grid=none,  axis lines=none]
     	\addplot+[mark=none,domain=0:60,  samples=15,color=olive,style={ultra thick},style=dashed] 
		{-sin(3*x)}; 
   \end{polaraxis}
 \end{tikzpicture}}} \nonumber \\ & = & \Tr (L C_6 C_5 C_6^{\sf T} C_5^\dagger)+\Tr (C_6 L C_5 C_6^{\sf T} C_5^\dagger).
 \nonumber 
\eea }
$\bullet$ Order-6:
{\allowdisplaybreaks \ytableausetup{smalltableaux}
\bea
J^+_{l^2 {c}^2} &:=& \hspace{-0.0cm}{\raisebox{-1.5em}{\begin{tikzpicture}
  \begin{polaraxis}[grid=none,  axis lines=none]
     	\addplot+[mark=none,domain=0:90,  samples=15,color=pink,style={ultra thick}] 
		{sin(2*x)}; 
   \end{polaraxis}
   \begin{polaraxis}[grid=none,  axis lines=none]
     	\addplot+[mark=none,domain=90:180,  samples=15,color=pink,style={ultra thick}] 
		{sin(2*x)}; 
   \end{polaraxis}
   \begin{polaraxis}[grid=none,  axis lines=none]
     	\addplot+[mark=none,domain=180:270,  samples=15,color=darkgreen,style={ultra thick},style=dashed] 
		{sin(2*x)}; 
   \end{polaraxis}
     \begin{polaraxis}[grid=none,  axis lines=none]
     	\addplot+[mark=none,domain=270:360,  samples=15,color=darkgreen,style={ultra thick},style=dashed] 
		{sin(2*x)}; 
   \end{polaraxis}
 \end{tikzpicture}}}=\Tr(L^2 |C_5|^4), 
 \nonumber \\
J^+_{{c}^3} &:=&\hspace{-0.1cm}{\raisebox{-1.1em}{\begin{tikzpicture}
 \begin{polaraxis}[grid=none,  axis lines=none]
     	\addplot+[mark=none,domain=0:60,  samples=15,color=darkgreen,style={ultra thick},style=dashed] 
		{-sin(3*x)}; 
   \end{polaraxis}
    \begin{polaraxis}[grid=none,  axis lines=none]
     	\addplot+[mark=none,domain=60:120,  samples=15,color=darkgreen,style={ultra thick},style=dashed] 
		{-sin(3*x)}; 
   \end{polaraxis}
       \begin{polaraxis}[grid=none,  axis lines=none]
     	\addplot+[mark=none,domain=120:180,  samples=15,color=darkgreen,style={ultra thick},style=dashed] 
		{-sin(3*x)}; 
   \end{polaraxis}
 \end{tikzpicture}}} = \Tr (|C_5|^6),
 \nonumber \\
 J^+_{l^1{lk_2}^1} &:=&  \hspace{-0.1cm}{\raisebox{-1.1em}{\begin{tikzpicture}
 \begin{polaraxis}[grid=none,  axis lines=none]
     	\addplot+[mark=none,domain=60:120,  samples=15,color=pink,style={ultra thick}] 
		{-sin(3*x)}; 
   \end{polaraxis}
    \begin{polaraxis}[grid=none,  axis lines=none]
     	\addplot+[mark=none,domain=120:180,  samples=15,color=pink,style={ultra thick}] 
		{-sin(3*x)}; 
   \end{polaraxis}
       \begin{polaraxis}[grid=none,  axis lines=none]
     	\addplot+[mark=none,domain=0:60,  samples=15,color=purple,style={ultra thick},style=dashed] 
		{-sin(3*x)}; 
   \end{polaraxis}
 \end{tikzpicture}}}=\Tr(L^2 C_5 {L^{\sf T}}^2 C_5^\dagger), 
 \nonumber \\
 J^+_{c_6^2 {c}^2}  &:=& \hspace{-0.0cm}{\raisebox{-1.5em}{\begin{tikzpicture}
  \begin{polaraxis}[grid=none,  axis lines=none]
     	\addplot+[mark=none,domain=0:90,  samples=15,color=lime,style={ultra thick}] 
		{sin(2*x)}; 
   \end{polaraxis}
   \begin{polaraxis}[grid=none,  axis lines=none]
     	\addplot+[mark=none,domain=90:180,  samples=15,color=lime,style={ultra thick}] 
		{sin(2*x)}; 
   \end{polaraxis}
   \begin{polaraxis}[grid=none,  axis lines=none]
     	\addplot+[mark=none,domain=180:270,  samples=15,color=darkgreen,style={ultra thick},style=dashed] 
		{sin(2*x)}; 
   \end{polaraxis}
     \begin{polaraxis}[grid=none,  axis lines=none]
     	\addplot+[mark=none,domain=270:360,  samples=15,color=darkgreen,style={ultra thick},style=dashed] 
		{sin(2*x)}; 
   \end{polaraxis}
 \end{tikzpicture}}}=\Tr(C_6^2 |C_5|^4), 
 \nonumber \\
  J^+_{c_6^1{c_6h_2}^1} &:=&  \hspace{-0.1cm}{\raisebox{-1.1em}{\begin{tikzpicture}
 \begin{polaraxis}[grid=none,  axis lines=none]
     	\addplot+[mark=none,domain=60:120,  samples=15,color=lime,style={ultra thick}] 
		{-sin(3*x)}; 
   \end{polaraxis}
    \begin{polaraxis}[grid=none,  axis lines=none]
     	\addplot+[mark=none,domain=120:180,  samples=15,color=lime,style={ultra thick}] 
		{-sin(3*x)}; 
   \end{polaraxis}
       \begin{polaraxis}[grid=none,  axis lines=none]
     	\addplot+[mark=none,domain=0:60,  samples=15,color=violet,style={ultra thick},style=dashed] 
		{-sin(3*x)}; 
   \end{polaraxis}
 \end{tikzpicture}}}=\Tr(C_6^2 C_5 {C_6^{\sf T}}^2 C_5^\dagger), 
 \nonumber \\
  J^+_{c_6^1{c_6k_2}^1}  &:=& \hspace{-0.1cm}{\raisebox{-1.1em}{\begin{tikzpicture}
 \begin{polaraxis}[grid=none,  axis lines=none]
     	\addplot+[mark=none,domain=60:120,  samples=15,color=lime,style={ultra thick}] 
		{-sin(3*x)}; 
   \end{polaraxis}
    \begin{polaraxis}[grid=none,  axis lines=none]
     	\addplot+[mark=none,domain=120:180,  samples=15,color=lime,style={ultra thick}] 
		{-sin(3*x)}; 
   \end{polaraxis}
       \begin{polaraxis}[grid=none,  axis lines=none]
     	\addplot+[mark=none,domain=0:60,  samples=15,color=purple,style={ultra thick},style=dashed] 
		{-sin(3*x)}; 
   \end{polaraxis}
 \end{tikzpicture}}}=\Tr(C_6^2 C_5 {L^{\sf T}}^2 C_5^\dagger), 
 \nonumber \\ 
 J^+_{{lk_1}^1c^1} &:=&  \hspace{-0.1cm}{\raisebox{-1.1em}{\begin{tikzpicture}
 \begin{polaraxis}[grid=none,  axis lines=none]
     	\addplot+[mark=none,domain=0:60,  samples=15,color=darkgreen,style={ultra thick},style=dashed] 
		{-sin(3*x)}; 
   \end{polaraxis}
    \begin{polaraxis}[grid=none,  axis lines=none]
     	\addplot+[mark=none,domain=60:120,  samples=15,color=pink,style={ultra thick}] 
		{-sin(3*x)}; 
   \end{polaraxis}
       \begin{polaraxis}[grid=none,  axis lines=none]
     	\addplot+[mark=none,domain=120:180,  samples=15,color=orange,style={ultra thick},style=dashed] 
		{-sin(3*x)}; 
   \end{polaraxis}
 \end{tikzpicture}}}  \nonumber \\ &=& \Tr({L C_5 {L}^T{C_5}^\dagger |C_5|^2})
  \nonumber \\ & + &
 \Tr({ C_5  {L}^{\sf T}{C_5}^\dagger\, L |C_5|^2}),
 \nonumber \\ 
 J^+_{{lh_1}^1c^1} &:=&  \hspace{-0.1cm}{\raisebox{-1.1em}{\begin{tikzpicture}
 \begin{polaraxis}[grid=none,  axis lines=none]
     	\addplot+[mark=none,domain=0:60,  samples=15,color=darkgreen,style={ultra thick},style=dashed] 
		{-sin(3*x)}; 
   \end{polaraxis}
    \begin{polaraxis}[grid=none,  axis lines=none]
     	\addplot+[mark=none,domain=60:120,  samples=15,color=pink,style={ultra thick}] 
		{-sin(3*x)}; 
   \end{polaraxis}
       \begin{polaraxis}[grid=none,  axis lines=none]
     	\addplot+[mark=none,domain=120:180,  samples=15,color=olive,style={ultra thick},style=dashed] 
		{-sin(3*x)}; 
   \end{polaraxis}
 \end{tikzpicture}}}
 \nonumber \\ &=& \Tr({L C_5 {C_6}^T{C_5}^\dagger |C_5|^2})
   \nonumber \\ & + &
   \Tr({ C_5  {C_6}^{\sf T}{C_5}^\dagger\, L |C_5|^2})\, ,
 \nonumber \\ 
 J^+_{{c_6h_1}^1c^1} &:=&  \hspace{-0.1cm}{\raisebox{-1.1em}{\begin{tikzpicture}
 \begin{polaraxis}[grid=none,  axis lines=none]
     	\addplot+[mark=none,domain=0:60,  samples=15,color=darkgreen,style={ultra thick},style=dashed] 
		{-sin(3*x)}; 
   \end{polaraxis}
    \begin{polaraxis}[grid=none,  axis lines=none]
     	\addplot+[mark=none,domain=60:120,  samples=15,color=lime,style={ultra thick}] 
		{-sin(3*x)}; 
   \end{polaraxis}
       \begin{polaraxis}[grid=none,  axis lines=none]
     	\addplot+[mark=none,domain=120:180,  samples=15,color=olive,style={ultra thick},style=dashed] 
		{-sin(3*x)}; 
   \end{polaraxis}
 \end{tikzpicture}}}
  \nonumber \\ &=& \Tr({C_6 C_5 {C_6}^T{C_5}^\dagger |C_5|^2})
    \nonumber \\ & + &\Tr({ C_5  {C_6}^{\sf T}{C_5}^\dagger\, C_6 |C_5|^2})\, ,
 \nonumber \\
   J^+_{{c_6k_2}^1} &:=&   \hspace{-0.1cm}{\raisebox{-1.1em}{\begin{tikzpicture}
 \begin{polaraxis}[grid=none,  axis lines=none]
     	\addplot+[mark=none,domain=0:60,  samples=15,color=darkgreen,style={ultra thick},style=dashed] 
		{-sin(3*x)}; 
   \end{polaraxis}
    \begin{polaraxis}[grid=none,  axis lines=none]
     	\addplot+[mark=none,domain=60:120,  samples=15,color=lime,style={ultra thick}] 
		{-sin(3*x)}; 
   \end{polaraxis}
       \begin{polaraxis}[grid=none,  axis lines=none]
     	\addplot+[mark=none,domain=120:180,  samples=15,color=purple,style={ultra thick},style=dashed] 
		{-sin(3*x)}; 
   \end{polaraxis}
 \end{tikzpicture}}}  \nonumber \\ &=&  \Tr({L C_5 {{L}^{\sf T}}^2 {C_5}^\dagger C_6})
   \nonumber \\ & + &\Tr({ C_5  {{L}^{\sf T}}^2 {C_5}^\dagger\, L C_6 })\, ,
 \nonumber \\
  J^+_{{c_6h_2}^1} &:=&  \hspace{-0.1cm}{\raisebox{-1.1em}{\begin{tikzpicture}
 \begin{polaraxis}[grid=none,  axis lines=none]
     	\addplot+[mark=none,domain=0:60,  samples=15,color=darkgreen,style={ultra thick},style=dashed] 
		{-sin(3*x)}; 
   \end{polaraxis}
    \begin{polaraxis}[grid=none,  axis lines=none]
     	\addplot+[mark=none,domain=60:120,  samples=15,color=lime,style={ultra thick}] 
		{-sin(3*x)}; 
   \end{polaraxis}
       \begin{polaraxis}[grid=none,  axis lines=none]
     	\addplot+[mark=none,domain=120:180,  samples=15,color=violet,style={ultra thick},style=dashed] 
		{-sin(3*x)}; 
   \end{polaraxis}
 \end{tikzpicture}}}  \nonumber \\ &=&  \Tr({L C_5 {{C_6}^{\sf T}}^2 {C_5}^\dagger C_6})
   \nonumber \\ & + &\Tr({ C_5  {{C_6}^{\sf T}}^2 {C_5}^\dagger\, L C_6 })\, ,
 \nonumber \\
 J^+_{c_6^1 {lf^+_{11}}^1}&:=&  \hspace{-0.1cm}{\raisebox{-1.1em}{\begin{tikzpicture}
 \begin{polaraxis}[grid=none,  axis lines=none]
     	\addplot+[mark=none,domain=0:60,  samples=15,color=lime,style={ultra thick},style=dashed] 
		{-sin(3*x)}; 
   \end{polaraxis}
    \begin{polaraxis}[grid=none,  axis lines=none]
     	\addplot+[mark=none,domain=60:120,  samples=15,color=pink,style={ultra thick}] 
		{-sin(3*x)}; 
   \end{polaraxis}
       \begin{polaraxis}[grid=none,  axis lines=none]
     	\addplot+[mark=none,domain=120:180,  samples=15,color=blue,style={ultra thick},style=dashed] 
		{-sin(3*x)}; 
   \end{polaraxis}
 \end{tikzpicture}}}  \nonumber \\ &=& \Tr({L C_6 C_5 ({L^{\sf T} C_6^{\sf T}+ C_6^{\sf T}L^{\sf T}})   {C_5}^\dagger })
   \nonumber \\ &+&  \Tr({C_6 L C_5 ({L^{\sf T} C_6^{\sf T}+C_6^{\sf T}L^{\sf T}}) {C_5}^\dagger })\, ,
  \nonumber \\
     J^+_{c_6^1 {lf^-_{11}}^1}&:=&  \hspace{-0.1cm}{\raisebox{-1.1em}{\begin{tikzpicture}
 \begin{polaraxis}[grid=none,  axis lines=none]
     	\addplot+[mark=none,domain=0:60,  samples=15,color=lime,style={ultra thick},style=dashed] 
		{-sin(3*x)}; 
   \end{polaraxis}
    \begin{polaraxis}[grid=none,  axis lines=none]
     	\addplot+[mark=none,domain=60:120,  samples=15,color=pink,style={ultra thick}] 
		{-sin(3*x)}; 
   \end{polaraxis}
       \begin{polaraxis}[grid=none,  axis lines=none]
     	\addplot+[mark=none,domain=120:180,  samples=15,color=cyan,style={ultra thick},style=dashed] 
		{-sin(3*x)}; 
   \end{polaraxis}
 \end{tikzpicture}}}  \nonumber \\ &=&  \Tr({L C_6 C_5 ({L^{\sf T} C_6^{\sf T}- C_6^{\sf T}L^{\sf T}})   {C_5}^\dagger })
   \nonumber \\ &-&  \Tr({C_6 L C_5({L^{\sf T} C_6^{\sf T}- C_6^{\sf T}L^{\sf T}}) {C_5}^\dagger })\, ,
  \label{sec:d-5o8-3}
  \nonumber
\eea}
where the dashed petals in violet, purple, blue and cyan present order four  $K_2 \equiv C_5 {L^{\sf T}}^2 C_5^\dagger$, $H_2 \equiv C_5 {C_6^{\sf T}}^2 C_5^\dagger$ and $F_{11}^\pm\equiv C_5 ({L^{\sf T}}{C_6^{\sf T}} \;\pm\; {C_6^{\sf T}}{L^{\sf T}}) C_5^\dagger$.

In addition to the aforementioned invariants, there are CP-odd and joint invariants. The first CP-odd invariant appears at order four and comprises nine invariants. The first joint invariant emerges at order five, where, in combination with four CP-odd and basic invariants, one can account for 14 invariants. Similarly, at order six, along with the previous 13 invariants, there are an additional 20 CP-odd and joint invariants.
Note that, there are CP-odd $J^-$ relevant to each $J^+$ invariant containing two terms but for their subtractions. However, there is only one $J^-$ invariant corresponding to the two last relations. 

Until the fifth order, all invariants are of the basic type. Yet, from this order onward, joint invariants can also be recognised.

In the subsequent listing, we present CP-odd invariants, commencing from the lowest conceivable order exhibiting CP-violation.
{\allowdisplaybreaks \ytableausetup{smalltableaux}
\bea
J^-_{{lc_6}^1c^1} &:=&  \hspace{-0.1cm}{\raisebox{-1.1em}{\begin{tikzpicture}
 \begin{polaraxis}[grid=none,  axis lines=none]
     	\addplot+[mark=none,domain=120:180,  samples=15,color=lime,style={ultra thick}] 
		{-sin(3*x)}; 
   \end{polaraxis}
    \begin{polaraxis}[grid=none,  axis lines=none]
     	\addplot+[mark=none,domain=60:120,  samples=15,color=pink,style={ultra thick}] 
		{-sin(3*x)}; 
   \end{polaraxis}
       \begin{polaraxis}[grid=none,  axis lines=none]
     	\addplot+[mark=none,domain=0:60,  samples=15,color=darkgreen,style={ultra thick},style=dashed] 
		{-sin(3*x)}; 
   \end{polaraxis}
 \end{tikzpicture}}}
 \nonumber \\ &=&\Tr(L C_6 |C_5|^2 )\;- \Tr( L |C_5|^2 C_6 ),
  \\
 J^-_{l^1{lc_6}^1c^1} &:=& \hspace{-0.1cm}{\raisebox{-1.1em}{\begin{tikzpicture}
 \begin{polaraxis}[grid=none,  axis lines=none]
     	\addplot+[mark=none,domain=0:180,  samples=15,color=pink,style={ultra thick}] 
		{sin(2*x)}; 
   \end{polaraxis}
    \begin{polaraxis}[grid=none,  axis lines=none]
     	\addplot+[mark=none,domain=180:270,  samples=15,color=lime,style={ultra thick}] 
		{ sin(2*x)}; 
   \end{polaraxis}
       \begin{polaraxis}[grid=none,  axis lines=none]
     	\addplot+[mark=none,domain=270:360,  samples=15,color=darkgreen,style={ultra thick},style=dashed] 
		{ sin(2*x)}; 
   \end{polaraxis}
 \end{tikzpicture}}}  \nonumber \\ &=&\Tr(L^2 C_6 |C_5|^2 )- \Tr( L^2 |C_5|^2 C_6 ),
  \\
  J^-_{c_6^1{lc_6}^1c^1} &:=&   \hspace{-0.1cm}{\raisebox{-1.1em}{\begin{tikzpicture}
 \begin{polaraxis}[grid=none,  axis lines=none]
     	\addplot+[mark=none,domain=0:90,  samples=15,color=pink,style={ultra thick}] 
		{sin(2*x)}; 
   \end{polaraxis}
    \begin{polaraxis}[grid=none,  axis lines=none]
     	\addplot+[mark=none,domain=90:270,  samples=15,color=lime,style={ultra thick}] 
		{ sin(2*x)}; 
   \end{polaraxis}
       \begin{polaraxis}[grid=none,  axis lines=none]
     	\addplot+[mark=none,domain=270:360,  samples=15,color=darkgreen,style={ultra thick},style=dashed] 
		{ sin(2*x)}; 
   \end{polaraxis}
 \end{tikzpicture}}} \nonumber \\ &=&\Tr(L C_6^2 |C_5|^2 ) - \Tr( L |C_5|^2 C_6^2 ),
  \\
 J^-_{l^1{c_6k_1}^1}&:=&  \hspace{-0.1cm}{\raisebox{-1.1em}{\begin{tikzpicture} 
 \begin{polaraxis}[grid=none,  axis lines=none]
     	\addplot+[mark=none,domain=60:120,  samples=15,color=pink,style={ultra thick}] 
		{-sin(3*x)}; 
   \end{polaraxis}
    \begin{polaraxis}[grid=none,  axis lines=none]
     	\addplot+[mark=none,domain=120:180,  samples=15,color=lime,style={ultra thick}] 
		{-sin(3*x)}; 
   \end{polaraxis}
       \begin{polaraxis}[grid=none,  axis lines=none]
     	\addplot+[mark=none,domain=0:60,  samples=15,color=orange,style={ultra thick},style=dashed] 
		{-sin(3*x)}; 
   \end{polaraxis}
 \end{tikzpicture}}} \nonumber \\ &=& \Tr (L C_6 C_5 L^{\sf T} C_5^\dagger)
  \nonumber \\ &-& \Tr (C_6 L C_5 L^{\sf T} C_5^\dagger),
 \\ 
  J^-_{l^1{c_6h_1}^1}&:=& \hspace{-0.1cm}{\raisebox{-1.1em}{\begin{tikzpicture}
 \begin{polaraxis}[grid=none,  axis lines=none]
     	\addplot+[mark=none,domain=60:120,  samples=15,color=pink,style={ultra thick}] 
		{-sin(3*x)}; 
   \end{polaraxis}
    \begin{polaraxis}[grid=none,  axis lines=none]
     	\addplot+[mark=none,domain=120:180,  samples=15,color=lime,style={ultra thick}] 
		{-sin(3*x)}; 
   \end{polaraxis}
       \begin{polaraxis}[grid=none,  axis lines=none]
     	\addplot+[mark=none,domain=0:60,  samples=15,color=olive,style={ultra thick},style=dashed] 
		{-sin(3*x)}; 
   \end{polaraxis}
 \end{tikzpicture}}} \nonumber \\ &=& \Tr (L C_6 C_5 C_6^{\sf T} C_5^\dagger)
  \nonumber \\ &-&\Tr (C_6 L C_5 C_6^{\sf T} C_5^\dagger),
\\
 J^-_{{lc_6}^1c^2} &:=&   \hspace{-0.1cm}{\raisebox{-1.1em}{\begin{tikzpicture}
 \begin{polaraxis}[grid=none,  axis lines=none]
     	\addplot+[mark=none,domain=0:90,  samples=15,color=pink,style={ultra thick}] 
		{sin(2*x)}; 
   \end{polaraxis}
    \begin{polaraxis}[grid=none,  axis lines=none]
     	\addplot+[mark=none,domain=90:180,  samples=15,color=lime,style={ultra thick}] 
		{sin(2*x)}; 
   \end{polaraxis}
       \begin{polaraxis}[grid=none,  axis lines=none]
     	\addplot+[mark=none,domain=180:360,  samples=15,color=darkgreen,style={ultra thick},style=dashed] 
		{sin(2*x)}; 
   \end{polaraxis}
 \end{tikzpicture}}}  \nonumber \\ &=&\Tr(L C_6 |C_5|^4 )- \Tr( L |C_5|^4 C_6)\, .
\eea }
From the aforementioned invariants, it can be observed that CPV occurs \textit{if and only if} ${\rm Im}J^-\neq 0$. It is worth noting that when constructing invariants of higher dimensions than eight, certain invariants involving higher orders of $C_{5,6}$ can be omitted.

In cases involving degeneracies, the assessment of higher-order CP-violating invariants becomes crucial. Alongside $J^-$, which aligns with the petals and relations for $J^+$ (but with reversed signs between terms) as illustrated in the fundamental invariant at the sixth order, a comprehensive list of additional higher-order CP-violating invariants is provided in Appendix~\ref{app:CPO}.

\section{Comparison with Hilbert Series}\label{sec:HS}

The Hilbert series, along with its associated PL, are powerful tools in the quest for understanding invariants in theories, but they do have limitations. These methods offer an algebraic enumeration of invariants, providing a generative function for constructing them; however, they may not always elucidate the underlying structure and interrelationships among these invariants, especially in the context of complex field theories and higher-dimensional operators. One intrinsic limitation is that the Hilbert series can obscure the physical interpretation of the invariants it enumerates. It tends to provide a raw count without distinguishing between invariants of different physical significance or between redundant and essential invariants. This can lead to an overestimation of the independent invariants when relations among invariants are present. 

In light of these aspects, while the Hilbert series and PL remain invaluable in the field of invariant theory, there is a compelling need for complementary approaches that can provide a more direct and physically intuitive grasp of invariants. Such is the promise of the newly established ring-diagram technique, which, by manifesting the structure of invariants through a visual and topological framework, can potentially overcome some of the shortcomings of the purely algebraic methods, providing a more nuanced and complete understanding of the invariant landscape in various theories.

As demonstrated in Section \ref{sec:SSMEFT}, we have outlined the structure, number, and CP properties of invariants up to the sixth order in the $\nu$SMEFT framework with operators 5 and 6. An interesting exercise is to compare the magnitude of these invariants with predictions from the Hilbert series~\cite{Yu:2021cco,Yu:2022ttm}. We start by defining the character functions in terms of ${\bf 3}$ and ${\bf 3}^*$ as $z_1 + z_2 + z_3$ and $z_1^{-1} + z_2^{-1} + z_3^{-1}$, respectively, where $z_{1,2}$ and $z_3$ are coordinates on the maximal torus of $U(3)$. The character functions of the flavor invariants, $\left\{L \equiv Y_eY_e^\dagger, C_5, C_6 \right\}$, analogous to Table \ref{tab:tabdim56}, are expressed as:
\begin{eqnarray}
\chi_{l,6} (z_1, z_2, z_3) &=& (z_1 + z_2 + z_3)(z_1^{-1} + z_2^{-1} + z_3^{-1}),\nonumber\\
&& 
\\
\chi_5 (z_1, z_2, z_3) &=& z_1^2 + z_2^2 + z_3^2 + z_1 z_2 + z_1 z_3 + z_2 z_3 +z_1^{-2} 
\nonumber\\&+&  z_2^{-2} +z_3^{-2} + z_1^{-1}z_2^{-1} + z_1^{-1}z_3^{-1} + z_2^{-1}z_3^{-1}.
\nonumber\\
\end{eqnarray}
Now, by incorporating the above relations into the Plethystic Exponential (PE) function and the Hilbert series for the $\nu$SMEFT framework with operators of dimensions 5 and 6, denoted as ${\cal H}_6(q)$, and considering the $3\times3$ matrices (detailed in Appendix~\ref{app:DHS}), we define the PL as follows:
\begin{eqnarray}
	\label{eq:PL}
	{\rm PL}\left[{\cal H}_6(q)\right]&=&2q+4q^2+6q^3+9q^4+14q^5+33q^6+44q^7
 \nonumber \\
&+&72q^8+74q^9+21q^{10}-{\cal O}\left(q ^{11}\right).
\end{eqnarray}
From this, the magnitude of the invariants can be extracted, although the underlying structure and properties of the invariants remain concealed. Comparing this with our finding in Table~\ref{tab:termination} the order of the leading negative term matches with the formulation (Orders $>10 \equiv$ LO of CP odd + the maximum between HO basic and LO joint) obtained using the ring-diagram. For sanity check, we compare these findings with more EFT models listed in Table~\ref{tab:taball}. The summary of the lowest order of CP-odd due to matrix properties (not because of their complex nature), the highest order of basic and the lowest order of joint invariants for SM, SMEFT and $\nu$SMEFT are given. 
\begin{table}[H]
\centering
\footnotesize 
\setlength{\tabcolsep}{3pt} 
\begin{tabular}{|l l|c c c |c|}
 \hline
	No & Models & LO\,CP-odd & HO\,basic & LO\,joint & Excl. \\\hline
	1 & SM & 6 & 4 & 6 & $\geq$ 12 \\\hline
	2 & SMEFT-dim$2n$ & 3 & 4 & 4 & $\geq$7 \\\hline
	3 & $\nu$-SMEFT-dim5 & 6 & 6 & 7 & $\geq$13 \\\hline
	4 & $\nu$-SMEFT-dim5,6 & 4 & 6 & 5 & $>$10 \\\hline
	5 & $\nu$-SMEFT-dim5,7 & 5 & 6 & 6 & $\geq$11 \\\hline
\end{tabular}
\caption{\it The Lowest Order (LO) of CP-odd invariants due to matrix properties (not their complex nature) are given in the first column. The Highest Order (HO) of basic and the LO of joint invariants are detailed in the subsequent columns. Finally, the last column shows the order in which invariants terminate.}
\label{tab:taball}
\end{table}
The PL equations corresponding to the models listed in Table~\ref{tab:taball} are as follows:
{\allowdisplaybreaks
\begin{eqnarray}
\small 
{\rm PL}_{\rm 1}\left[{\cal H}(q)\right] &=& 2 q + 3 q^2 + 4 q^3 + q^4 + q^6 
\nonumber \\
&&- {\cal O}\left(q ^{12}\right)\;,
\\
{\rm PL}_{2}[{\cal H}(q)] &=& 4 q + 10 q^2 + 24 q^3 + 35 q^4 + 56 q^5+ 60 q^6 \nonumber \\
&& -\mathcal{O}(q^7)\;,
\label{PL-dim-6}
\\
{\rm PL}_{3}\left[{\cal H}(q)\right] &=& q + 2 q^2 + 2 q^3 + 3 q^4 + 2 q^5 + 5 q^6 + 2 q^7\nonumber \\
&&  + 5 q^8  +4 q^9 +  5 q^{10} + 2 q^{11} 
\nonumber \\
&&-{\cal O}\left(q ^{13}\right)\;,
\\
{\rm PL}_{4}\left[{\cal H}_6(q)\right] &=& 2q+4q^2+6q^3+9q^4+14q^5+33q^6+44q^7 \nonumber \\
&& +72q^8+74q^9+21q^{10}
\nonumber \\
&&-{\cal O}\left(q ^{11}\right) \;,
\\
{\rm PL}_{5}\left[{\cal H}(q)\right] &=& q + 5 q^2 + 5 q^3 + 17 q^4 + 20 q^5  + 82 q^7\nonumber \\
&& + 175 q^8 + 231 q^9 + 199 q^{10}
\nonumber \\
&& -{\cal O}\left(q ^{11}\right)\;.
\end{eqnarray}}

The exclusion order in the last column in Table \ref{tab:taball} is aligned with the negative leading term in the above PL related to SM and EFT models.

\section{Conclusions}\label{sec:generic}
We present an automatic and novel
methodology for explicitly classifying CP invariants. We utilise fundamental blocks, constructed as
orthogonal trivial singlets via newly established ring-diagrams, and employ the Cayley-Hamilton theorem for
precise identification.
We delineate an automatic mechanism for differentiating between CP-odd and -even basis invariants.

Remarkably, this approach not only provides a complementary way of identifying invariants but also provides a new explanation for details in the traditional Hilbert-Poincar\'e series and its Plethystic logarithm (PL) concerning basis invariant, joint and negative leading numbers in PL.

Utilising this method, we successfully identify the complete set of SM invariants through the consideration of two fundamental matrices.
We have demonstrated the application of this approach in the context of the seesaw Standard Model Effective Field Theory ($\nu$SMEFT), focusing on operators of dimensions 5 and 6. Our technique is developed as a general methodology, adaptable to SMEFT for incorporating higher-order operators up to dimension-$2n$. It is also extendible to $\nu$SMEFT with operators of dimensions 5, 6, and 7. These extensions are subjects of our forthcoming publications.

Additionally, future studies will detail the application of this method in constructing invariants from high-rank tensors and complex structures. Moreover, we will explore its applications within the frameworks of multi-Higgs doublet models ($n$HDMs) and $n$HDM-Effective Field Theory ($n$HDM-EFT)~\cite{Darvishi:2019dbh,Darvishi:2022wnd,Birch-Sykes:2020btk}.

\subsection*{ACKNOWLEDGMENTS} 
This work is supported by the National Science
Foundation of China under Grants No.~12022514,
No.~12375099 and No.~12047503, and the National Key
Research and Development Program of China Grant
No.~2020YFC2201501, and No.~2021YFA0718304 and
by the CAS President's International Fellowship Initiative
(PIFI) grant. The work of ND is also supported by STFC
under the Grant No.~ST/T006749/1. 

\appendix

\section{IDENTIFYING HIGHER-ORDER
CP-ODD INVARIANTS}\label{app:CPO}
In this Appendix, we provide a list of higher-order CP-odd invariants in the framework of $\nu$SMEFT with operators of dim-5 and dim-6.
In Section~\ref{sec:SSMEFT}, we have shown the list of basic CP-even and six of the lowest order CP-odd invariants in this framework. Following the structure of ring-diagram higher-order CP-odd invariants starting from order-6 are:
{\allowdisplaybreaks \ytableausetup{smalltableaux}
\bea
J^-_{lc_6^2{c}^2}&:=&  \hspace{-0.1cm}{\raisebox{-1.1em}{\begin{tikzpicture}
 \begin{polaraxis}[grid=none,  axis lines=none]
     	\addplot+[mark=none,domain=0:36,  samples=15,color=pink,style={ultra thick} ] 
		{sin(5*x)}; 
   \end{polaraxis}
   \begin{polaraxis}[grid=none,  axis lines=none]
     	\addplot+[mark=none,domain=36:72,  samples=15,color=lime,style={ultra thick} ] 
		{sin(5*x)}; 
   \end{polaraxis}
    \begin{polaraxis}[grid=none,  axis lines=none]
     	\addplot+[mark=none,domain=72:108,  samples=15,color=pink,style={ultra thick} ] 
		{sin(5*x)}; 
   \end{polaraxis}
       \begin{polaraxis}[grid=none,  axis lines=none]
     	\addplot+[mark=none,domain=108:144,  samples=15,color=lime,style={ultra thick} ] 
		{sin(5*x)}; 
   \end{polaraxis}
       \begin{polaraxis}[grid=none,  axis lines=none]
     	\addplot+[mark=none,domain=144:180,  samples=15,color=darkgreen,style={ultra thick,dashed}] 
		{sin(5*x)}; 
   \end{polaraxis}
 \end{tikzpicture}}}  \nonumber \\ &=& \Tr (L^2C_6^2|C_5|^2)-\Tr (C_6^2L^2|C_5|^2),
\nonumber \\
 J^-_{l^2 {lc_6}^1c^1}&:=&  \hspace{-0.1cm}
{\raisebox{-1.4em}{\begin{tikzpicture}
  \begin{polaraxis}[grid=none,  axis lines=none]
     	\addplot+[mark=none,domain=0:36,  samples=15,color=pink,style={ultra thick}] 
		{sin(5*x)}; 
   \end{polaraxis}
   \begin{polaraxis}[grid=none,  axis lines=none]
     	\addplot+[mark=none,domain=36:72,  samples=15,color=pink,style={ultra thick}] 
		{sin(5*x)}; 
   \end{polaraxis}
   \begin{polaraxis}[grid=none,  axis lines=none]
     	\addplot+[mark=none,domain=72:108,  samples=15,color=pink,style={ultra thick}] 
		{sin(5*x)}; 
   \end{polaraxis}
     \begin{polaraxis}[grid=none,  axis lines=none]
     	\addplot+[mark=none,domain=108:144,  samples=15,color=lime,style={ultra thick} ] 
		{sin(5*x)}; 
   \end{polaraxis}
 \begin{polaraxis}[grid=none, axis lines=none]
     	\addplot+[mark=none,domain=144:180,  samples=15,color=darkgreen,style={ultra thick,dashed} ] 
		{sin(5*x)}; 
\end{polaraxis}
 \end{tikzpicture}}} \nonumber \\ &=&\Tr({L^2 C_6 L |C_5|^2})-\Tr(L^2 |C_5|^2 L C_6),
\nonumber \\
 J^-_{c_6^2 {lc_6}^1c^1}&:=&  \hspace{-0.1cm}
{\raisebox{-1.4em}{\begin{tikzpicture}
  \begin{polaraxis}[grid=none,  axis lines=none]
     	\addplot+[mark=none,domain=0:36,  samples=15,color=lime,style={ultra thick}] 
		{sin(5*x)}; 
   \end{polaraxis}
   \begin{polaraxis}[grid=none,  axis lines=none]
     	\addplot+[mark=none,domain=36:72,  samples=15,color=lime,style={ultra thick}] 
		{sin(5*x)}; 
   \end{polaraxis}
   \begin{polaraxis}[grid=none,  axis lines=none]
     	\addplot+[mark=none,domain=72:108,  samples=15,color=pink,style={ultra thick}] 
		{sin(5*x)}; 
   \end{polaraxis}
     \begin{polaraxis}[grid=none,  axis lines=none]
     	\addplot+[mark=none,domain=108:144,  samples=15,color=darkgreen,style={ultra thick,dashed} ] 
		{sin(5*x)}; 
   \end{polaraxis}
 \begin{polaraxis}[grid=none, axis lines=none]
     	\addplot+[mark=none,domain=144:180,  samples=15,color=lime,style={ultra thick} ] 
		{sin(5*x)}; 
\end{polaraxis}
 \end{tikzpicture}}} \nonumber \\ &=& \Tr({C_6^2 L C_6 |C_5|^2})-\Tr(C_6^2 |C_5|^2 C_6 L),
\nonumber \\
J^-_{{lc_6}^3}&:=&  \hspace{0.1cm}{\raisebox{-1.5em}{\begin{tikzpicture}
 \begin{polaraxis}[grid=none,  axis lines=none]
     	\addplot+[mark=none,domain=0:270,  samples=30,color=pink,style={ultra thick}] 
		{-sin(3*x)}; 
   \end{polaraxis}
   \begin{polaraxis}[grid=none,  axis lines=none]
     	\addplot+[mark=none,domain=0:270,  samples=30,color=lime,style={ultra thick}] 
		{sin(3*x)}; 
   \end{polaraxis}
 \end{tikzpicture}}} \nonumber \\ &=&\Tr({L^2 C_6^2 L C_6})-\Tr(C_6^2 L^2 C_6 L),
\nonumber \\
 J^-_{l^2 {c_6k_1}^1} &:=&  \hspace{-0.1cm}{\raisebox{-1.1em}{\begin{tikzpicture}
 \begin{polaraxis}[grid=none,  axis lines=none]
     	\addplot+[mark=none,domain=0:36,  samples=15,color=pink,style={ultra thick} ] 
		{sin(5*x)}; 
   \end{polaraxis}
   \begin{polaraxis}[grid=none,  axis lines=none]
     	\addplot+[mark=none,domain=36:72,  samples=15,color=lime,style={ultra thick} ] 
		{sin(5*x)}; 
   \end{polaraxis}
    \begin{polaraxis}[grid=none,  axis lines=none]
     	\addplot+[mark=none,domain=72:108,  samples=15,color=pink,style={ultra thick} ] 
		{sin(5*x)}; 
   \end{polaraxis}
       \begin{polaraxis}[grid=none,  axis lines=none]
     	\addplot+[mark=none,domain=108:144,  samples=15,color=lime,style={ultra thick} ] 
		{sin(5*x)}; 
   \end{polaraxis}
       \begin{polaraxis}[grid=none,  axis lines=none]
     	\addplot+[mark=none,domain=144:180,  samples=15,color=orange,style={ultra thick},style=dashed] 
		{sin(5*x)}; 
   \end{polaraxis}
 \end{tikzpicture}}} \nonumber \\ &=& \Tr({L^2 C_5 L^{\sf T} {C_5}^\dagger C_6^2})
  \nonumber \\ &-&\Tr({ C_5  L^{\sf T}{C_5}^\dagger\, L^2  C_6^2}),
\label{sec:d-56o6}
\nonumber \\
J^-_{l^2 {c_6h_1}^1} &:=&  \hspace{-0.1cm}{\raisebox{-1.1em}{\begin{tikzpicture}
 \begin{polaraxis}[grid=none,  axis lines=none]
     	\addplot+[mark=none,domain=0:36,  samples=15,color=pink,style={ultra thick} ] 
		{sin(5*x)}; 
   \end{polaraxis}
   \begin{polaraxis}[grid=none,  axis lines=none]
     	\addplot+[mark=none,domain=36:72,  samples=15,color=lime,style={ultra thick} ] 
		{sin(5*x)}; 
   \end{polaraxis}
    \begin{polaraxis}[grid=none,  axis lines=none]
     	\addplot+[mark=none,domain=72:108,  samples=15,color=pink,style={ultra thick} ] 
		{sin(5*x)}; 
   \end{polaraxis}
       \begin{polaraxis}[grid=none,  axis lines=none]
     	\addplot+[mark=none,domain=108:144,  samples=15,color=lime,style={ultra thick} ] 
		{sin(5*x)}; 
   \end{polaraxis}
       \begin{polaraxis}[grid=none,  axis lines=none]
     	\addplot+[mark=none,domain=144:180,  samples=15,color=olive,style={ultra thick},style=dashed] 
		{sin(5*x)}; 
   \end{polaraxis}
 \end{tikzpicture}}} \nonumber \\ &=&\Tr({L^2 C_5 C_6^{\sf T} {C_5}^\dagger C_6^2})
 \nonumber \\ &-&\Tr({ C_5  C_6^{\sf T}{C_5}^\dagger\, L^2  C_6^2}),
\label{sec:d-56o6}
\nonumber \\
J^-_{c_6^2 {lk_1}^1} &:=&  \hspace{-0.1cm}{\raisebox{-1.1em}{\begin{tikzpicture}
 \begin{polaraxis}[grid=none,  axis lines=none]
     	\addplot+[mark=none,domain=0:90,  samples=15,color=lime,style={ultra thick}] 
		{-sin(2*x)}; 
   \end{polaraxis}
     \begin{polaraxis}[grid=none,  axis lines=none]
     	\addplot+[mark=none,domain=180:270,  samples=15,color=pink,style={ultra thick}] 
		{-sin(2*x)}; 
   \end{polaraxis}
    \begin{polaraxis}[grid=none,  axis lines=none]
     	\addplot+[mark=none,domain=270:360,  samples=15,color=lime,style={ultra thick}] 
		{-sin(2*x)}; 
   \end{polaraxis}
       \begin{polaraxis}[grid=none,  axis lines=none]
     	\addplot+[mark=none,domain=90:180,  samples=15,color=orange,style={ultra thick},style=dashed] 
		{-sin(2*x)}; 
   \end{polaraxis}
 \end{tikzpicture}}} \nonumber \\ &=& \Tr({L C_6^2 C_5  L^{\sf T} {C_5}^\dagger})
 \nonumber \\ &-&\Tr({ L C_5  L^{\sf T}{C_5}^\dagger\,  C_6^2}),
\label{sec:d-56po6} \nonumber \\
J^-_{c_6^2 {lh_1}^1} &:=&  \hspace{-0.1cm}{\raisebox{-1.1em}{\begin{tikzpicture}
 \begin{polaraxis}[grid=none,  axis lines=none]
     	\addplot+[mark=none,domain=0:90,  samples=15,color=lime,style={ultra thick}] 
		{-sin(2*x)}; 
   \end{polaraxis}
     \begin{polaraxis}[grid=none,  axis lines=none]
     	\addplot+[mark=none,domain=180:270,  samples=15,color=pink,style={ultra thick}] 
		{-sin(2*x)}; 
   \end{polaraxis}
    \begin{polaraxis}[grid=none,  axis lines=none]
     	\addplot+[mark=none,domain=270:360,  samples=15,color=lime,style={ultra thick}] 
		{-sin(2*x)}; 
   \end{polaraxis}
       \begin{polaraxis}[grid=none,  axis lines=none]
     	\addplot+[mark=none,domain=90:180,  samples=15,color=olive,style={ultra thick},style=dashed] 
		{-sin(2*x)}; 
   \end{polaraxis}
 \end{tikzpicture}}} \nonumber \\ &=&\Tr({L C_6^2 C_5 C_6^{\sf T} {C_5}^\dagger C_6})
  \nonumber \\ &-&\Tr({ L C_5  C_6^{\sf T}{C_5}^\dagger\,  C_6^2}),
\label{sec:d-56po6} 
\nonumber \\
 J^-_{{lk_1}^1c^1} &:=&  \hspace{-0.1cm}{\raisebox{-1.1em}{\begin{tikzpicture}
 \begin{polaraxis}[grid=none,  axis lines=none]
     	\addplot+[mark=none,domain=0:60,  samples=15,color=orange,style={ultra thick},style=dashed] 
		{-sin(3*x)}; 
   \end{polaraxis}
    \begin{polaraxis}[grid=none,  axis lines=none]
     	\addplot+[mark=none,domain=60:120,  samples=15,color=pink,style={ultra thick}] 
		{-sin(3*x)}; 
   \end{polaraxis}
       \begin{polaraxis}[grid=none,  axis lines=none]
     	\addplot+[mark=none,domain=120:180,  samples=15,color=darkgreen,style={ultra thick},style=dashed] 
		{-sin(3*x)}; 
   \end{polaraxis}
 \end{tikzpicture}}} \nonumber \\ &=&\Tr({L C_5 {L}^T{C_5}^\dagger |C_5|^2})
 \nonumber \\ &-&\Tr({ C_5  {L}^{\sf T}{C_5}^\dagger\, L |C_5|^2}),
\nonumber \\
 J^-_{l^1 {lk_1}^1c^1} &:=&  \hspace{-0.1cm}{\raisebox{-1.1em}{\begin{tikzpicture}
 \begin{polaraxis}[grid=none,  axis lines=none]
     	\addplot+[mark=none,domain=0:90,  samples=15,color=darkgreen,style={ultra thick},style=dashed] 
		{-sin(2*x)}; 
   \end{polaraxis}
    \begin{polaraxis}[grid=none,  axis lines=none]
     	\addplot+[mark=none,domain=180:360,  samples=15,color=pink,style={ultra thick}] 
		{-sin(2*x)}; 
   \end{polaraxis}
       \begin{polaraxis}[grid=none,  axis lines=none]
     	\addplot+[mark=none,domain=90:180,  samples=15,color=orange,style={ultra thick},style=dashed] 
		{-sin(2*x)}; 
   \end{polaraxis}
 \end{tikzpicture}}} \nonumber \\ &=&\Tr({L^2 C_5 L^{\sf T} {C_5}^\dagger |C_5|^2})
  \nonumber \\ &-&\Tr({ C_5  L^{\sf T}{C_5}^\dagger\, L^2  |C_5|^2}),
\label{sec:d-56o7} \nonumber \\
J^-_{ {lk_1}^1c^2} &:=&   \hspace{-0.1cm}{\raisebox{-1.1em}{\begin{tikzpicture}
 \begin{polaraxis}[grid=none,  axis lines=none]
     	\addplot+[mark=none,domain=0:90,  samples=15,color=darkgreen,style={ultra thick},style=dashed] 
		{-sin(2*x)}; 
   \end{polaraxis}
   \begin{polaraxis}[grid=none,  axis lines=none]
     	\addplot+[mark=none,domain=270:360,  samples=15,color=darkgreen,style={ultra thick},style=dashed] 
		{-sin(2*x)}; 
   \end{polaraxis}
    \begin{polaraxis}[grid=none,  axis lines=none]
     	\addplot+[mark=none,domain=180:270,  samples=15,color=pink,style={ultra thick}] 
		{-sin(2*x)}; 
   \end{polaraxis}
       \begin{polaraxis}[grid=none,  axis lines=none]
     	\addplot+[mark=none,domain=90:180,  samples=15,color=orange,style={ultra thick},style=dashed] 
		{-sin(2*x)}; 
   \end{polaraxis}
 \end{tikzpicture}}} \nonumber \\ &=&\Tr({L C_5 {L}^{\sf T}{C_5}^\dagger |C_5|^4 }) 
  \nonumber \\ &-& \Tr({ C_5  {L}^{\sf T}{C_5}^\dagger\, L  |C_5|^4} ),
 \nonumber \\
 J^-_{l^1 {lk_2}^1c^1}&:=&  \hspace{-0.1cm}{\raisebox{-1.1em}{\begin{tikzpicture}
 \begin{polaraxis}[grid=none,  axis lines=none]
     	\addplot+[mark=none,domain=0:90,  samples=15,color=darkgreen,style={ultra thick},style=dashed] 
		{-sin(2*x)}; 
   \end{polaraxis}
    \begin{polaraxis}[grid=none,  axis lines=none]
     	\addplot+[mark=none,domain=180:360,  samples=15,color=pink,style={ultra thick}] 
		{-sin(2*x)}; 
   \end{polaraxis}
       \begin{polaraxis}[grid=none,  axis lines=none]
     	\addplot+[mark=none,domain=90:180,  samples=15,color=purple,style={ultra thick},style=dashed] 
		{-sin(2*x)}; 
   \end{polaraxis}
 \end{tikzpicture}}} \nonumber \\ &=&\Tr({C_5  {L^{\sf T}}^2{C_5}^\dagger |C_5|^2  L^2 })  \nonumber \\ &-& \Tr({ C_5  {L^{\sf T}}^2 {C_5}^\dagger L^2 |C_5|^2 }),
 \nonumber \\
 J^-_{l^1 {lk_1}^1c^1}&:=&  \hspace{-0.1cm}
{\raisebox{-1.4em}{\begin{tikzpicture}
  \begin{polaraxis}[grid=none,  axis lines=none]
     	\addplot+[mark=none,domain=0:36,  samples=15,color=pink,style={ultra thick}] 
		{sin(5*x)}; 
   \end{polaraxis}
   \begin{polaraxis}[grid=none,  axis lines=none]
     	\addplot+[mark=none,domain=36:72,  samples=15,color=pink,style={ultra thick}] 
		{sin(5*x)}; 
   \end{polaraxis}
   \begin{polaraxis}[grid=none,  axis lines=none]
     	\addplot+[mark=none,domain=72:108,  samples=15,color=pink,style={ultra thick}] 
		{sin(5*x)}; 
   \end{polaraxis}
     \begin{polaraxis}[grid=none,  axis lines=none]
     	\addplot+[mark=none,domain=108:144,  samples=15,color=darkgreen,style={ultra thick},style=dashed] 
		{sin(5*x)}; 
   \end{polaraxis}
 \begin{polaraxis}[grid=none, axis lines=none]
     	\addplot+[mark=none,domain=144:180,  samples=15,color=orange,style={ultra thick},style=dashed] 
		{sin(5*x)}; 
\end{polaraxis}
 \end{tikzpicture}}} \nonumber \\ &=&\Tr({L^2 C_5  {L^{\sf T}} {C_5}^\dagger L |C_5|^2 })  \nonumber \\ &-& \Tr({L^2 |C_5|^2 L C_5  {L^{\sf T}} {C_5}^\dagger}).
  \nonumber
\eea}
\begin{widetext}
\section{HILBERT SERIES FOR $\nu$SMEFT FRAMEWORK WITH OPERATORS
\\ OF DIMENSIONS 5, 6}
\label{app:DHS}
In Section~\ref{sec:HS} we have shown the character functions of the flavor invariants, $\left\{L \equiv Y_eY_e^\dagger, C_5, C_6 \right\}$ can be given in the following forms
\begin{eqnarray}
\chi_{l,6} (z_1, z_2, z_3) &=& (z_1 + z_2 + z_3)(z_1^{-1} + z_2^{-1} + z_3^{-1}), \nonumber \\
\chi_5 (z_1, z_2, z_3) &=& z_1^2 + z_2^2 + z_3^2 + z_1 z_2 + z_1 z_3 + z_2 z_3 \nonumber\\
+z_1^{-2} &+& z_2^{-2} + z_3^{-2} + z_1^{-1}z_2^{-1} + z_1^{-1}z_3^{-1} + z_2^{-1}z_3^{-1}. \nonumber
\end{eqnarray}
Pertinent to the above character functions, the PE function can be expressed as:

\begin{eqnarray}
  \label{eq:PEdim56}
  {\rm PE}(z_1 ,z_2 ,z_3 &;&q ) \,=\, {\rm exp} \left(\sum_{k=1}^\infty\frac{\chi_l (z_1^k,z_2^k,z_3^k )q^k+\chi_5 (z_1^k,z_2^k,z_3^k )q^k+\chi_6 (z_1^k,z_2^k,z_3^k )q^k}{k} \right) \nonumber\\
  &=& \Big[(1-q ) ^6 (1-q z_1  z_2^{-1} ) ^2 (1-q z_2  z_1^{-1} ) ^2 (1-q z_1  z_3^{-1} ) ^2 (1-q z_3  z_1^{-1} ) ^2 (1-q z_2  z_3^{-1} ) ^2
 \nonumber\\
&&    (1-q z_3  z_2^{-1}) ^2  (1-q z_1^2 ) (1-q z_2^2) (1-q z_3^2 ) (1-q z_1 z_2) (1-q z_1 z_3)  (1-q z_2 z_3)
\nonumber\\
&&   
   (1-q z_1^{-2} ) (1-q z_2^{-2} ) (1-q z_3^{-2} ) (1-q z_1^{-1}z_2^{-1} )  (1-q z_1^{-1}z_3^{-1} ) (1-q z_2^{-1}z_3^{-1} )\Big] ^{-1} .
\end{eqnarray}

Thus Hilbert series in the $\nu$SMEFT with operators of dimensions 5 and 6 for the $3\times3$ matrices becomes

{\begin{eqnarray}
		{\cal H}_6 (q)&=&\frac{{ \cal N}_6 (q)}{{ \cal D}_6 (q)}\;=\;\int \Big[{\rm d}\mu\Big]_{\rm U (3)}  {\rm PE} (z_1 ,z_2 ,z_3 ;q )\nonumber\\
	&=& \frac{1}{ 6(2\pi {\rm i} )^3}\oint_{\left|z_1\right|=1}\oint_{\left|z_2\right|=1}\oint_{\left|z_3\right|=1}\Big[-\frac{(z_2-z_1 )^2 (z_3-z_1 )^2 (z_3-z_2 )^2}{z_1^2 z_2^2 z_3^2}\Big]{\rm PE} (z_1 ,z_2 ,z_3;q )\;.
\end{eqnarray}}

In the above relation the nominator ${\cal N}_6 (q)$ and the denominator ${\cal D}_6 (q)$ read
{\allowdisplaybreaks
\begin{eqnarray}
\label{eq:numerator2}
{\cal N}_6 (q) & = & 1 + 2 q^3 + 4 q^4 + 11 q^5 + 33 q^6 + 52 q^7  +  104 q^8 + 182 q^9 + 307 q^{10} + 495 q^{11} + 808 q^{12} \nonumber \\
& + & 1176 q^{13} + 1692 q^{14} + 2307 q^{15} + 2995 q^{16}  +  3736 q^{17} + 4546 q^{18} + 5246 q^{19} + 5902 q^{20} \nonumber \\
& + & 6401 q^{21} + 6632 q^{22} + 6632 q^{23} + 6401 q^{24} + 5902 q^{25} + 5246 q^{26} + 4546 q^{27} + 3736 q^{28} \nonumber \\
& + & 2995 q^{29} + 2307 q^{30} + 1692 q^{31} + 1176 q^{32} +808 q^{33} + 808 q^{33} + 495 q^{34} + 307 q^{35} + 182 q^{36} 
\nonumber \\
& + & 104 q^{37}  +52 q^{38} + 33 q^{39} + 11 q^{40} + 4 q^{41} + 2 q^{42} + q^{45},
\end{eqnarray}}
and 
\begin{eqnarray}
\label{eq:denominator}
{\cal D}_6 (q)&=&\left(1 - q\right)^2 \left(1 - q^2\right)^4 \left(1 - q^3\right)^4 \left(1 - q^4\right)^5  
 \left(1 - q^5\right)^3  \left(1 -  q^6\right)^3\;.
\end{eqnarray}
Finally, the associated PL for Hilbert series can be derived using the following relation:
\begin{eqnarray}
\label{eq:PL}
{\rm PL}\left[{\cal H}\left(q\right)\right] \equiv \sum_{k=1}^{\infty} \frac{\mu(k)}{k}\,{\rm ln}\left[{\cal H}\left(q^k\right)\right]\;,
\end{eqnarray}
where $\mu(k)$ is the M{\"o}bius function~\cite{Pouliot_1999}, defined as:
\begin{equation}
\label{eq:mob}
\mu(k) \equiv
\begin{cases}
0             &  \text{ repeated prime factors in $ k $ } \\
1              &  \text{ $ k = 1 $} \\
(-1)^n &  \text{product of $ n $ distinct primes in $k$}
\end{cases}\;.
\end{equation}
\end{widetext}

\bibliographystyle{apsrev4-1} 
\bibliography{Refs}
\clearpage

\end{document}